\documentclass[prd,twocolumn,showpacs,amssymb]{revtex4}

\usepackage{amsmath, amssymb, graphics}
\usepackage{longtable}
\usepackage{dcolumn}
\usepackage{latexsym} 
\usepackage{graphicx} 
\usepackage{subfigure} 
\usepackage{amsmath} 
\usepackage{supertabular} 
\usepackage{natbib} 
\usepackage[latin1]{inputenc} 
\usepackage[thinspace]{SIunits} 

\footnotesize
\newdimen\minuswidth    
\setbox0=\hbox{$-$}
\minuswidth=\wd0
\catcode`@=\active
\def@{\kern\minuswidth}
\newdimen\digitwidth    
\setbox0=\hbox{\rm0}
\digitwidth=\wd0
\catcode`!=\active
\def!{\kern\digitwidth}
\normalsize

\begin{document}

\title{Gravitational wave signal of the short rise fling of galactic run 
away pulsars \\ ACCEPTED FOR PUBLICATION IN JCAP 17/10/2008}

\author{Herman J. Mosquera Cuesta$^{1,2}$, Carlos A. Bonilla Quintero$^1$}

\affiliation{\mbox{$^1$Instituto de Cosmologia, Relatividade e Astrof\'\i sica 
(ICRA-BR),  Centro Brasileiro de Pesquisas F{\'\i}sicas} \\ 
\mbox{Rua Dr. Xavier Sigaud 150, CEP 22290-180, Urca Rio de Janeiro, RJ, Brazil}  } 

\newcommand{\setthebls}{
}
\setthebls

\begin{abstract}
Determination of pulsar parallaxes and proper motions addresses fundamental 
astrophysical open issues. Here, after scrutinizing the ATNF Catalog searching 
for pulsar distances and proper motions,  we verify that for an ATNF sample of 
212 Galactic run away pulsars (RAPs), which currently run across the Galaxy at 
very high speed and 
undergo large displacements, some gravitational-wave (GW) signals produced by 
such present accelerations appear to be detectable after calibration against the 
Advanced LIGO (LIGO II). Motivated by this insight, we address the issue of the 
pulsar kick at birth, or {\sl short rise fling} from a supernova explosion, by 
adapting the theory for emission of GW by ultrarelativistic sources to this case 
in which Lorentz factor is $\gamma \sim 1$. We show that during the short rise 
fling each run away pulsar (RAP) generates a GW signal with characteristic 
amplitude and frequency that makes it detectable by current GW interferometers. 
For a realistic analysis, an efficiency 
parameter is introduced to quantify the expenditure of the rise fling kinetic energy, 
which is estimated from the linear momentum conservation law applied to the 
supernova explosion that kicks out the pulsar. The remaining energy is supposed 
to be used to make the star to spin. Thus, a comparison with the spin of ATNF 
pulsars having velocity in the interval $400-500$ km s$^{-1}$ is performed. The 
resulting difference 
suggests that other mechanisms (like differential rotation, magnetic breaking 
or magneto-rotational instability) should dissipate part of that energy to 
produce the observed pulsar spin periods. Meanwhile, the kick phenomenon may 
also occur in globular and open star clusters at the formation or disruption 
of very short period compact binary systems wherein abrupt velocity and 
acceleration similar to those given to RAPs during the short rise fling 
can be imparted to each orbital partner. To better analyzing these cases, 
pulsar astrometry 
from micro- to nano-arsec scales might be of much help. In case of a 
supernova, the RAP GW signal could be a benchmark for the GW signal 
from the core collapse. 
\end{abstract}



\email{hermanjc@cbpf.br, herman@icra.it}

\date{\today}

\maketitle

\section{ Astrophysical Motivation}

The detection of gravitational waves (GW) is one of the greatest goals of 
today's relativistic astrophysics. Because of the great success of Einstein's 
general relativity (GR) in explaining the dynamical evolution of the binary 
pulsar PSR 1913+16; continuously observed over almost 30 years by Taylor and 
Hulse \cite{taylor2004}, most researchers in the field are confident that the 
very first GW signal to be detected must come from coalescing binary neutron 
star-neutron star or black hole-neutron star (NS-NS, BH-NS) systems. Ongoing 
LIGO science runs \cite{Abbott2007} for the first time have set firm upper limits 
on the rate of binary coalescences in our Galaxy: at the level of a few events per 
year. Meanwhile the model-dependent estimate of such a rate at cosmic distance scales 
varies around $10^{-5}$ \cite{postnov2006,kalogera2007,pacheco2006}. The expectation 
is enormous as well for an event like the supernova (SN) explosion. 

{ In this paper we focus on the radiation emitted when a pulsar is kicked out 
from the supernova remnant and is accelerated with very high speed, albeit 
non ultrarelativistic. Gravitational radiation is emitted whenever the nascent 
neutron star or pulsar changes its velocity, a condition that is fulfilled by 
the large thrust given to the nascent pulsar during the SN explosion, and also
due to the pulsar dragging against the SN ejecta. (Recall that the gravitational 
field generated by a motion with constant velocity is nonradiative). We shall 
show that each individual event of launching a nascent pulsar to drift or run 
across the Galaxy, after a SN explosion, generates 
a gravitational wave signal that can be detected by current and planned GW 
interferometers such LIGO I, LIGO II, VIRGO, GEO-600, or TAMA-300. Nonetheless, 
it is stressed from the very beginning that the strain of the GW signal that we 
estimate below is {\sl a lower bound}. The reason is that we are not taking into 
account the contribution to the GW amplitude, $h$, stemming from the effective 
acceleration that the pulsar receives at the kick at birth, which could be very 
large. Indeed, the distance over which the RAP is dragged against the supernova 
ejecta could also be astronomically large.}

At first glance, it appears that the detection of the GW signal from the short rise 
must have an event rate on the order of that for SN events in our galaxy, i. e. 
$\sim$ 1 per 
30-300 years. In principle such a rate has to be the same because the flinging 
of the pulsar could be just the aftermath of most supernova core collapses. However, 
in galactic globular clusters, where the frequency of formation and 
disruption of binary systems harboring compact stars is relatively high, the kick 
velocity of a pulsar partner at the binary disruption could be of the same magnitude 
as the one for galactic RAPs. Indeed, Vlemmings et al. \cite{pulsars-vlemmings2005} 
are pursuing a VLBI astrometry program to perform measurements of pulsar proper 
motions and parallaxes with spatial resolution of microarsecond. As an illustration 
of the performance of the high precision pulsar astrometry system they simulated 
the detection of the disruption of a binary system in the stellar cluster Cygnus 
Superbubble. This is a quite estimulating perspective for pulsar astrometry in the 
near future. 

Therefore, there is a potential to detect such a disruption or formation 
of a binary system through the GW signal emitted by the abrupt acceleration or sudden 
increase in velocity of a given pulsar. Observations in the range of micro- or 
nano-astrometry with planned satellites like GAIA and other instruments on Earth 
and space may help to identify such a phenomenon. An advanced technology useful for 
this purpose is currently operative or under planning (see Ref.\cite{pulsars-gaia} 
and references therein).



\begin{figure*}
\centering
\subfigure[\label{k=a} Pulsar number (vertical axis) vs. distance (kpc) 
to Earth (horizontal axis) histogram ] {
\includegraphics[height=3.0in,width=3.0in]{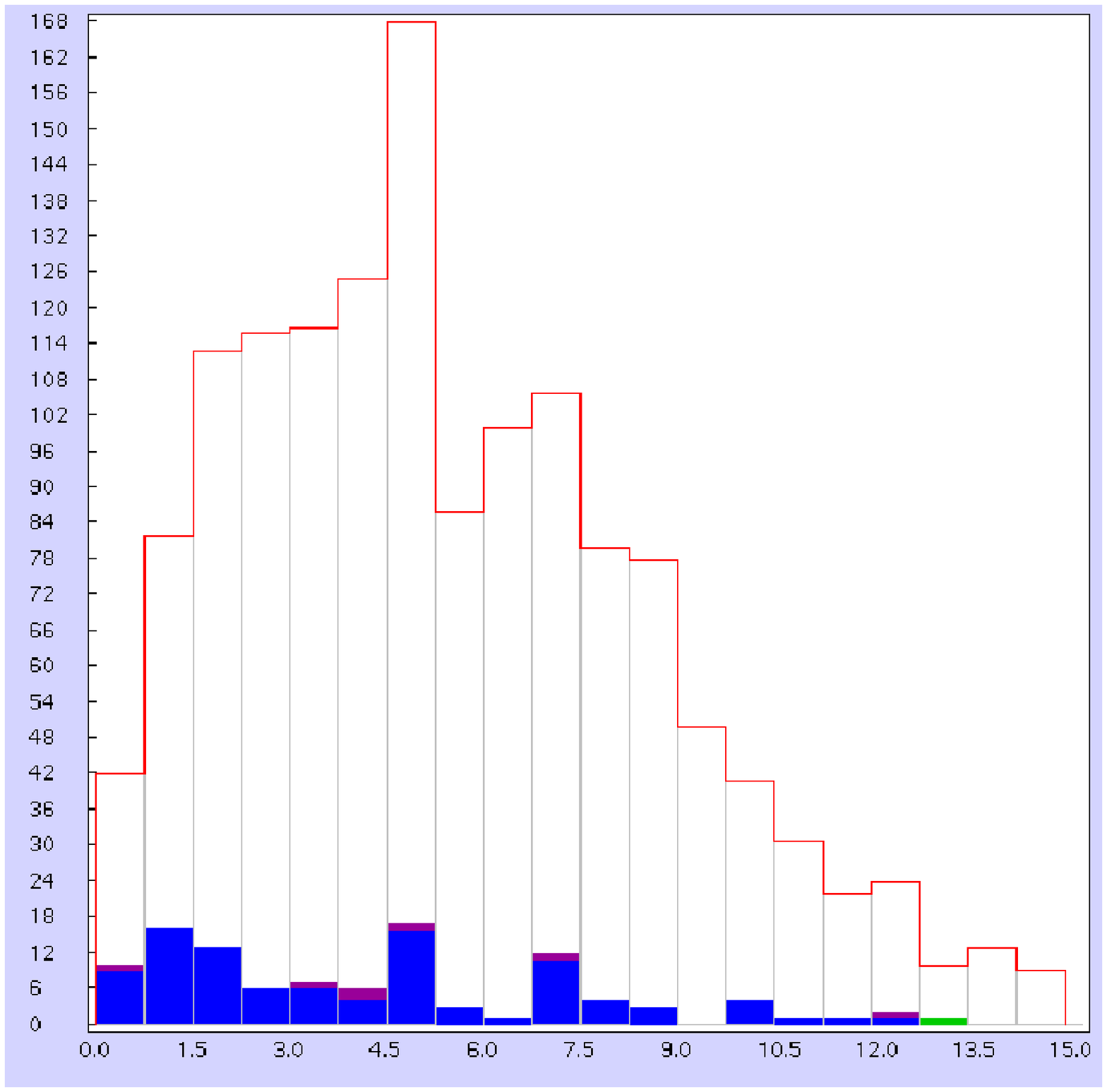}}
\hskip 0.5truecm
\subfigure[\label{k=b} Pulsar number vs. Distance distribution of 
212 pulsars ]{
\includegraphics[height=3.0in,width=3.0in]{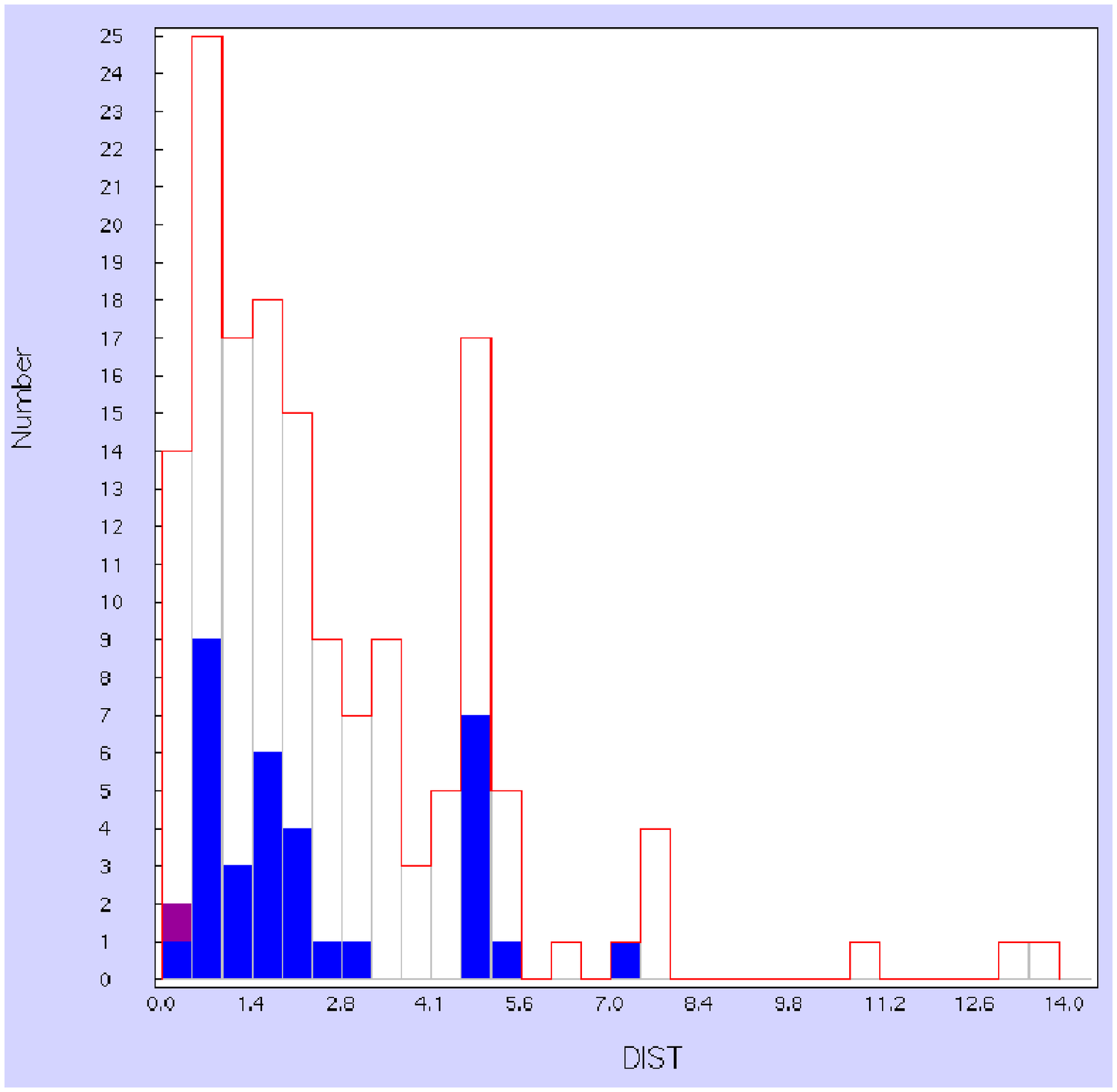}}
\hskip 0.5truecm
\subfigure[\label{PSR-VELOC-DIST} Velocity vs. Distance distribution of 212 
ATNF pulsars having both parameters measured ] {
\includegraphics[height=2.750in,width=2.750in]{distrib_veloc.eps} }
\caption{Data taken from ATNF Pulsar Catalog} 
\label{GW-HISTOGRAM-ATNF}
\end{figure*}


The physical reason for advocating in favor of this novel possibility
for GW detection is based on a number of results on the dynamics and 
kinematics of galactic RAPs  and also on GR, as is shown next. 
 
\begin{itemize}
\item 1) In a seminal paper, Spruit  and
Phinney \cite{phinney-spruit98}  demonstrated that regardless of the
actual physical engine driving the natal (instantaneous) pulsar kicks,
that mechanism can only act upon the pulsar over a very restricted time
scale, $\Delta T^{\rm kick}_{\rm max}$, on the order of $\simeq 0.32$~s. 
This is a fundamental piece of the dynamics of observed pulsars that 
seems to be overlooked by workers in the field. 

\item 2) As shown in the next Section, 
pulsar surveys have shown that the largest ($V^{\rm max}_\star$), smaller
($V^{\rm min}_\star$) and average ($V^{\rm ave}_\star$) spatial (3-D)
velocity of cataloged pulsars are 5000 km s$^{-1}$, 84 km~s$^{-1}$, and
450-500 km~s$^{-1}$, respectively \cite{lorimer93, arzoumian2002, Hobbs2005} 
(see Fig.\ref{PSR-VELOC-DIST}). That is, $V_\star$ changes by more than an 
order of magnitude. 

\item 3) Observations of NSs in binary systems have allowed to 
estimate their masses with very high accuracy. The  current values of this 
physical property gather around 1.4 M$_\odot$ \cite{chakra-thorsett}.  

\item 4) According to GR the acceleration of a mass ($M_\star$) to a
relativistic velocity ($V_\star$); i. e. energy, should generate a GW
pulse, the nature of which has a very distinctive imprint: the so-called
gravitational wave memory \cite{braginsky-thorne83,ori2001,sago04}, which 
makes the amplitude of this GW signal (the GW strain) not to fade away by 
the end of the traction phase (the kick), but rather leaves it variable 
in time as much as an energy source remains active in the source. 
\end{itemize}


\section{Galactic RAPs and sample selection from ATNF Catalog}

Run away pulsars (RAPs) are observed to drift along the Galaxy with velocities ranging 
from 90 km s$^{-1}$ to 4000 km s$^{-1}$ (Australian Telescope National Facility (ATNF) 
Catalog \cite{atnf-manchester2004}, \cite{PMBS}). 
Since  most stars in our galaxy  are observed to drift with average velocities  
of $15-30$ km~s$^{-1}$, it has been suggested that such large velocities should  
be given to the proto-neutron star  at its birth. The kick velocity imparted 
to a just-born neutron star during the rise time is a fundamental piece for 
understanding the physics of the core collapse supernovae, to have an insight on 
the sources of asymmetries during the gravitational collapse and on the emission 
of gravitational waves during this process. There seems to be a consensus that the 
core collapse asymmetry driving the kick velocity of a neutron star at birth may 
have origin in a variety of mechanisms \cite{dong-lai2001,dl2004}. 

\begin{figure}
\centering
\subfigure[ GW signals from pulsar acceleration and 
gamma-ray burst. Distribution (with the viewing angle) of the GW signal produced 
during the early acceleration phase of a pulsar (dashed--green line) with
$V^{\rm ave}_\star = 450$~km~s$^{-1}$, and distance = 10~kpc.
Comparison with the GW signal from a gamma-ray burst (solid--red line),
as a function of the jet angle to the line-of-sight, with
parameters $\langle E \rangle = 10^{51}$~erg, $\gamma = 100$, and distance =
1~Mpc.] {
\includegraphics[height=3.0in,width=3.0in]{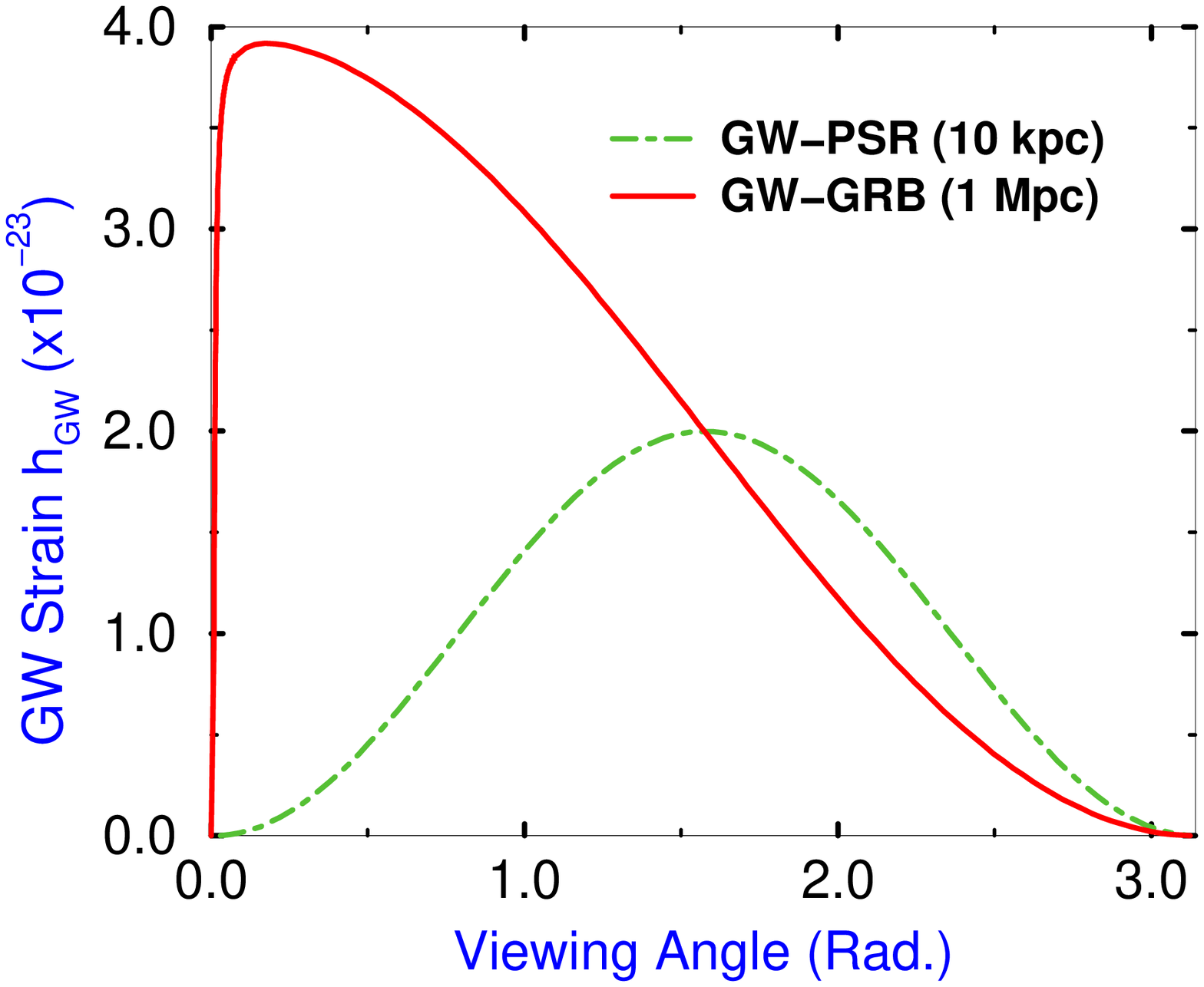}}
\caption{Angular distribution of the GW signals from pulsar acceleration and 
gamma-ray burst} 
\label{GW-PSR-GRB-SIGNAL}
\end{figure}


Although the mechanism responsible for the initial impulses has not been  properly 
identified yet, a very strong constraint on the timescale over which it could act 
on the nascent neutron star (NS) was put forward in Ref.\cite{phinney-spruit98}. 
Whichever this thruster  engine might be it cannot push the  star over a timescale 
longer than $\Delta T^{\rm kick}_{\rm max} \simeq 0.32$~s. Combining this with the 
NS mass, which piles-up around 1.4~M $_\odot$ \cite{chakra-thorsett}, and the above  
average velocity, implies that a huge power should be expended by the thruster during 
the pulsar early acceleration phase, and hereby a gravitational-wave (GW) burst with 
``memory" (understood as a steady time variation of the metric perturbation) must be 
released during this time, according to general relativity \cite{braginsky-thorne83}. 


The present velocity distribution (see Figs.\ref{GW-HISTOGRAM-ATNF}, 
\ref{PSR-VELOC-DIST}) is constructed based on the analysis of the proper motion of 
radio pulsars in our Galaxy after estimating their distances. Pulsar distances are 
estimated by using annual parallax, H$_{\rm I}$ absorption, or associations with 
globular clusters or supernova remnants, and from dispersion measures and galactic 
electron density models. 

This astronomy field has become a vivid research enterprise since the pioneer works 
of Lyne and Lorimer (LL94) \cite{lorimer93,lorimer94}. However, the current view on the 
distribution of kick velocities is far from the original one they found through pulsar 
surveys fifteen years ago. LL94 showed that pulsars appear to run around the Galaxy 
(and its halo) with  mean spatial (3-D) velocities $\sim 450-500$ km~s$^{-1}$ 
\cite{lorimer93}, but freeway (``cannon-ball'') pulsars with velocities  as high 
as 5000 km~s$^{-1}$ and slow-down millisecond pulsars with speeds of 80 km s$^{-1}$ 
wander also around the Galaxy \cite{lorimer94}. Meanwhile, the situation changed 
dramatically with the pulsar velocity survey performed by Arzoumian, Chernoff and Cordes 
(ACC02) \cite{arzoumian2002}, who found that such a speed distribution appeared 
to be bimodal indicating one distribution mode  around 90 km s$^{-1}$ and another 
near 500 km s$^{-1}$. However, this dynamic research field has shown, through new 
studies on pulsar proper motions, that the situation is far from having been settled.
For instance, Hobbs et al. \cite{Hobbs2005} have recently challenged the ACC02 
findings, and have shown that their studies lead to assert that there is no 
evidence of a bimodal distribution. Instead they suggest that mean 3-Dim pulsar
birth velocity is 400$\pm 40$ km~s$^{-1}$, and that the distribution appears to 
be Maxwellian with 1-Dim rms $\sigma = 265$ km~s$^{-1}$, showing also a continuum 
distribution from low to high (present) run away speeds,  being the last ones  
the tail of the 
velocity distribution. Hence, looking back on the surveys by LL94, the study by 
Hobbs et al. \cite{Hobbs2005} clearly agrees with the mean velocity of 450 $\pm 
90$ km s$^{-1}$ found by LL94 \cite{lorimer93,lorimer94}.

\subsection{Sample selection from ATNF database}

In the late times the number of pulsars has increased considerably \cite{Hobbs2005}. 
Although various researchers have kept update catalogues, since then; in general, 
these have been neither completed nor very accessible. The ATNF database is an 
effort among an australian scientific community and some foreing partner institutions. 
The purpose is to compile the largest amount of information about pulsars, which is 
subsequently made available to every researcher all over the world.

The ATNF catalogue \cite{Hobbs2005} registers all published rotation-powered pulsars, 
including those detected only at high energies. It also records Anomalous X-ray Pulsars 
AXPs and Soft Gamma-ray Repeaters (SGRs) for which coherent pulsations have been detected. 
However, it excludes accretion-powered pulsars such as Her X-1 and the recently 
discovered X-ray millisecond pulsars, for instance SAX J1808.4-3658 \cite{Wijnands1998}.

The ATNF catalogue \cite{Hobbs2005} can be accessed in a number of different ways.  The 
simplest one is from a web interface \footnote{http://www.atnf.csiro.au/research/pulsar/psrcat} 
allowing listing of the most commonly used pulsar parameters, together with the uncertainties 
and information on the references. Several options for {\sl tabular output} format are provided. 
Currently, a total of 67 ``{\sl predefined parameters}'' are available. A tool is provided 
for plotting of parameter distributions, either as two-dimensional plots or as histograms. 
Zoom tools and interactive identification of plotted points are provided. Custom parameters 
can be defined by combining parameters in expressions using mathematical operators and 
functions and these can be either listed or plotted.  Finally,  the sample of pulsars 
listed or plotted can be limited by logical conditions on parameters (see Table-\ref{212-atnf-psrs}), 
pulsar name (including wild-card names) or distance from a nominated position. These operational 
tools are described in more detail below and links are provided within the web interface 
to relevant documentation \cite{Hobbs2005}.

In this particular case, we select from ATNF, by clicking on the box to the left of the 
``parameter label'' to select a couple of pulsar parameters: DIST and VTRANS (see Fig.\ref{PSR-VELOC-DIST}). 
The first parameter is the best estimate of the pulsar distance in kpc. The second parameter 
is the transverse velocity, given in km s$^{-1}$,  based on DIST estimates. By default 
the results will be sorted according to the pulsars' J2000 names in ascending alphabetical 
order. However, sorting is possible on any parameter by typing the parameter label in the 
``{\sl sort on field text box}'' and selecting whether the sort should be in ascending (default) 
or descending order. However, to obtain as much information as possible some careful
procedures must be followed. Because of this, we selected the listing fashion ``{\sl 
short format without errors}''. The short format is used to provide a condensed summary 
of the pulsars parameters (see Table-\ref{periodos}). Finally, we selected the option 
``{\sl header}'', that produces a table with header information at the top and selected 
the output option ``{\sl Table}'' (see Table-\ref{212-atnf-psrs}). 
Besides, it is possible to display functions of the pulsar parameters as a graph or as 
a histogram. For a normal (x-y) graph, the values to plot are defined as regular 
expressions into the ``{\sl x, y-axis text box}'', and the axes of the graph can be displayed 
linearly or logarithmically. The expressions may also contain custom-defined variables or 
externally defined variables.

\section{\label{psr-galactic-dynamics} Present Galactic pulsar dynamics hints at
detectability of GW signals from the short rise fling of RAPs in supernovae}

Next we discuss the prospective for the short rise fling of Galactic RAPs in a 
supernova to generate detectable GW signals. The analysis starts from the observed 
kinematic state of RAPs, and goes back to their kicks at birth. To such an approach 
we assume that before the kick the star velocity is $V^0_\star = 0$ 
\cite{cordes-chernoff98}.

First, let us notice that one can combine the first three pieces discussed 
above together with the extremely restrictive character of the thrust timescale 
($\Delta T^{\rm kick}_{\rm max}$), the magnitude of the largest velocity 
($V^{\rm max}_\star$), and the instantaneous nature of the kick ($V^0_\star 
= 0$). In so doing, one can then conclude that whichever the driving engine 
might be, it should transfer to the star at the kick at birth an energy as large 
as $\langle \Delta E \rangle \sim 10^{51}$~erg. (Recall that the NS masses 
are piled-up around 1.4~M$_\odot$). Besides, keep in mind that according to 
Newton's second law: $F_\star = M_\star a_\star$, the star's inertia depends 
directly on its rest mass $M_\star$. Thus, putting together these ingredients 
one can conclude that the effective acceleration, $a_\star$, at the kick at 
birth of each RAP should be the same.\footnote{Notice that here we are 
assuming that the measured velocities are  the very initial ones. That is,  
we are neglecting the effects of the gravitational potential  of the galactic
central distribution of mass, the bulge, which can affect the presently
observed kinematics and energetics, especially, of  young pulsars born
near the galactic plane \cite{cordes-chernoff98}.} In other words, because 
pulsar velocities vary by more than an order of magnitude, whilst the inertia 
(mass) of each neutron star is essentially the same (within an uncertainty no 
greater then a factor of 2), one concludes that the acceleration at the kick 
at birth for each of the pulsars has definitely to be the same. In all our 
analysis below we will use this key feature. 

Further, as the kick is an ``instantaneous action'', then the Newtonian kinematics 
states that the largest velocity\footnote{We stress that even if larger 
velocities were found in forthcoming pulsar surveys \cite{cordes-chernoff98}, 
this would not change significantly the picture here presented.} should be
attained after the longest acceleration phase. That is; over the
longest  traction time $t^{\rm kick}_\star$. Therefore, the RAP 
kinematics during the short rise fling follows the law

\begin{equation}
V^{\rm fin}_\star(t) = V^0_\star + a_\star \cdot t_\star^{\rm kick} , 
\label{max-veloc}
\end{equation}

where $V^{\rm fin}_\star(t)$ defines the final (at the end of the kick)
velocity of a given RAP, which is reached after  a total impulsion time
$t_\star^{\rm kick} $. Consequently, if one assumes for the sake of 
simplicity that the RAP transit through the interstellar space does 
not modify its starting state of drifting, 
then one concludes that the largest (presently observed velocity $V^{\rm 
max}_\star \simeq 5000$ km~s$^{-1}$) must have been reached after the 
elapsing of the longest timescale permitted by the kick mechanism: $\Delta 
T^{\rm kick}_{\rm max} \simeq 0.32$~s \cite{phinney-spruit98}.

Thus, after collecting all the pieces on the RAPs physics described above  
one can infer the characteristic acceleration received by a typical RAP. 
To this goal on can take the largest present velocity given by Refs.
\cite{lorimer93, lorimer94, arzoumian2002, Hobbs2005} (see Fig.-\ref{PSR-VELOC-DIST}), 
and the constraint on the maximum traction timescale for the kick at 
birth\cite{phinney-spruit98}. That acceleration then reads

\begin{equation}
a_\star = \frac{V^{\rm max}_\star}{\Delta T^{\rm kick}_{\rm max}} \sim 
10^4~\rm km~s^{-2} . 
\label{accel}
\end{equation}

\subsection{Estimates of present $h_c$, $f_{\rm GW}$ GW characteristics 
hint at prospective detection during RAPs short rise fling in supernovae}

The  above analysis suggests that the overall time scale for acceleration 
of any given RAP to its present velocity (under the assumption stated above) 
depends exclusively on the total energy with which it was thrusted. The actual 
amount of energy expended to thrust the star certainly has something to do with 
the nature of the driving mechanism. Hence, the GW dynamics to be discussed 
below, combined with observations of Galactic RAPs, may help to enlighten the 
path to unravel the actual thruster. Next we provide order of magnitude 
estimates for both GW amplitude and frequency based on parameters of RAPs 
currently cataloged. (In the next Section a rigurous derivation of these GW 
characteristics for the RAP kick at birth is presented).

\subsubsection{Frequency estimate}

From Eqs.(\ref{max-veloc})-(\ref{accel}) one can see that the traction time
scale for any given pulsar is strictly directly proportional to its present 
spatial velocity. This is a conclusion of basic importance. That is, the RAP current 
kinematics must retain information about what happened during its short rise 
fling in the supernova. Thus, the frequency of the GW burst emitted during the 
early impulsion 
of a pulsar in a supernova explosion can be assumed, in a very simplified 
picture, as being inversely related to the total acceleration time \cite{sago04}, 
that is,

\begin{equation}
f^{\rm kick}_{\rm GW} \simeq \frac{1}{t_\star^{\rm kick} } \; .
\label{frecuencia}
\end{equation}

\begin{figure}[tbh]
\vskip 1.0truecm
\centering
\includegraphics[height=3.0in,width=3.0in]{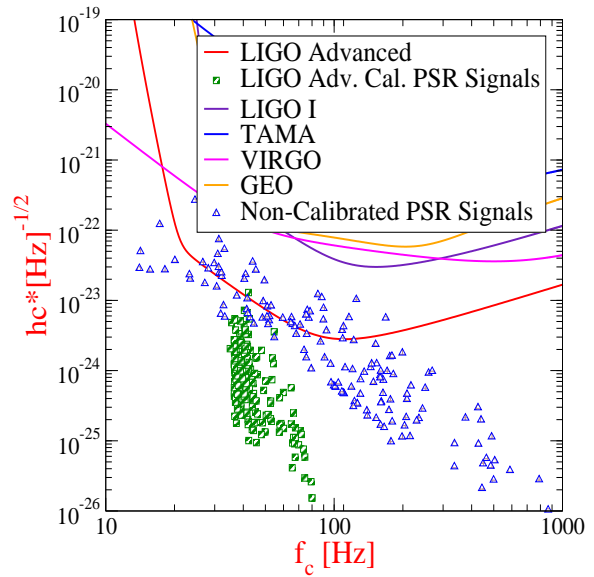}
\caption{{\it  Color Online. } LIGO I, LIGO Advanced, VIRGO, GEO-600, TAMA-300
strain sensitivities and the $h_{c}$ and $f_{c}$ characteristics of the GW 
signal produced by each of the 212 pulsars calibrated with respect to LIGO 
Advanced (green squares), and the same sample without calibration (blue 
triangles). The effect of the calibration procedure is evident.}
\label{approximate-sinal}
\end{figure}

As an  example, let us take the values quoted above for $V^{\rm max}_\star$,
$V^{\rm ave}_\star$, and  $V^{\rm min}_\star$. In these cases, one obtains 
the GW frequencies $f^{\rm min}_{\rm GW} = 3.1$~Hz, $f^{\rm  ave}_{\rm GW} =
34.7$~Hz, and $f^{\rm max}_{\rm GW} = 186.0$~Hz, respectively. It is
illustrative to compare these frequencies with those provided in 
Figs.-(\ref{L2efficiency-a6}, \ref{L2efficiency}, \ref{L1EFFICIENCY-a4}, 
\ref{EFF=CALIB-LIGO-I}) below. As shown next, the last ones would be detectable 
by present-day laser-beam interferometers whenever their associate strains 
(computed below) are large enough (see Fig.\ref{approximate-sinal} for the 
estimates of $h$ and $f_{\rm GW}$ for the 212 ATNF RAPs sample).

\subsubsection{Amplitude estimate}

To have an order of magnitude estimate of the GW amplitude, $h$, from the
present kinematic state of a galactic RAP one can use the general relativistic 
quadrupole formula

\begin{equation}
h \simeq \frac{4 G}{c^{4}} \frac{ E_{k} }{r}\; ,
\label{amplitude-approx}
\end{equation}

where $r$ is the pulsar distance as provided by the ATNF Catalog, and $E_{k} = \frac{1}{2} 
M_\star V^2_\star$ is the RAP kinetic energy. 

Therefore, if one combines the theoretical prediction provided by Eq.(\ref{frecuencia}) 
with this approximate GW amplitude, one obtains the plot presented in 
Fig.(\ref{approximate-sinal}). The estimates were performed for a sample 
of 212 pulsares selected from the Australian Telescope National Facility 
(ATNF) catalog \cite{atnf-manchester2004}.\footnote{As above-mentioned the 
ATNF catalog contains data on pulsar velocities, $V_\star$, distances, 
$r_\star$ and orientations $(\theta, \phi)$ angles with respect to the 
line of sight.} 
Notice that Eq.(\ref{amplitude-approx}) states that whenever the RAP 
velocity is time-varying the signals, whose representative points appear 
above the strain sensitivity limit of Advanced LIGO, would be detectable 
by this observatory. For a RAP with no time-varying velocity, the signal 
would be a direct-current-like (DC) signal that no detector might observe. 
(Regarding the above estimates, keep in mind that at least the pulsar 
motion around the Galactic center (the bulge) should be centripetally 
accelerated). 

In Fig.-\ref{approximate-sinal} blue 
triangles represent GW signals computed following this back-of-the-envelope 
technique. Green squares are the obtained after calibrating the approximated 
signals against the strain sensitivity of Advanced LIGO. Clearly only a few RAPs 
would be detectable if their GW signals were described in such an approximate 
fashion. Moreover, this graph gives us a clear hint at the proper characteristics 
of the GW signal that would be emitted by a run away pulsar during its short rise 
fling in the supernova that makes it to drift across the galaxy. These order of 
magnitude estimates suggest that a RAP that was kicked out from the supernova 
ejecta with higher speed and acceleration
would produce a clearly detectable GW signal. This perspective motivated us to study 
the problem using a more rigurous physical description: the theory for the emission 
of gravitational wave signals by ultrarelativistic sources, a GW signal which has 
memory. This is done in the next Section.



\begin{longtable*}{cccccccc}
\caption{\label{212-atnf-psrs}Full pulsar sample taken from ATNF.
JName is the pulsar name based on J2000 coordinates, RAJ right ascension (J2000) 
in hh:mm:ss.s, DECJ declination (J2000) in +dd:mm:ss, PMRA  proper motion 
in right ascension in mas yr$^{-1}$ units, PMDEC  proper motion in declination 
in mas yr$^{-1}$ units, DIST best estimate of the pulsar distance in kpc units, 
VTRANS transverse velocity - based on DIST in km s$^{-1}$} \\
\hline \hline
\#    & PSR[J]  & RAJ   & DECJ     & PMRA     &  PMDEC & DIST  & VTRANS  \\
{  }  & {Name  }  & (hms) & (dms) & (mas/yr) & (mas/yr) &  (kpc)  & (km/s)\\
\hline 
\hline
\endfirsthead
\caption{Full sample taken from ATNF Catalog.}\\
\hline \hline 
\#    & PSR[J]  & RAJ   & DECJ     & PMRA     &  PMDEC & DIST  & VTRANS  \\
{  }  & {Name  }  & (hms) & (dms) & (mas/yr) & (mas/yr) &  (kpc)  & (km/s)\\
\hline
\hline
\endhead  
1 & J0014+4746  & 00:14:17.7 & +47:46:33.4  & 19.300  &  -19.700   &  1.82 &  237.964  \\
2 & J0024-7204C & 00:23:50.3 & -72:04:31.4  & 5.200  &   -3.400  &    4.80  &  141.384  \\
3 & J0024-7204D & 00:24:13.8 & -72:04:43.8  & 3.800   &  -2.400   &   4.80  &  102.278  \\
4 & J0024-7204E & 00:24:11.1 & -72:05:20.1  & 6.200  &   -2.700   &   4.80  &  153.889  \\
5 & J0024-7204F & 00:24:03.8 & -72:04:42.8  & 4.600  &   -2.900  &    4.80  &  123.747  \\
6 & J0024-7204G & 00:24:07.9 & -72:04:39.6  & 4.200  &  -3.700  &    4.80  &  127.376  \\
7 & J0024-7204H & 00:24:06.7 & -72:04:06.7  & 5.100   & -3.600  &   4.80  &  142.060  \\
8 & J0024-7204I & 00:24:07.9 & -72:04:39.6  & 4.900  &   -3.900  &    4.80 &  142.515  \\
9 & J0024-7204J & 00:23:59.4 & -72:03:58.7  & 5.350   &  -3.420   &   4.80  &  144.498  \\
10 & J0024-7204N & 00:24:09.1 & -72:04:28.8 & 6.800  &   -1.100 &     4.80  &  156.757  \\
11 & J0024-7204O & 00:24:04.6 & -72:04:53.7 & 4.600  &   -2.900  &    4.80  &  123.747  \\
12 & J0024-7204U & 00:24:09.8 & -72:03:59.6 & 5.100   &  -4.500  &    4.80  &  154.779  \\
13 & J0030+0451  & 00:30:27.4 & +04:51:39.7 &  C    &     C  &    0.30  &    8.246  \\
14 &J0034-0534  & 00:34:21.8 & -05:34:36.6  &  C    &     C   &   0.98   & 146.280  \\
15 & J0040+5716 & 00:40:32.3 & +57:16:24.9  &  C   &      C   &   4.48 &   161.755  \\
16 & J0117+5914 & 01:17:38.6 & +59:14:38.3  &  C   &       C  &    2.14  &  186.526  \\
17 & J0134-2937 & 01:34:18.6 & -29:37:16.9  &  C    &      C   &   1.78  &  189.641  \\
18 & J0139+5814 & 01:39:19.7  &+58:14:31.7  & -21.000  &   -5.000   &   2.89  &  295.772  \\
19 & J0147+5922 & 01:47:44.6 & +59:22:03.2  &  C    &      C   &   1.91  &   96.870  \\
20 & J0151-0635 & 01:51:22.7 & -06:35:02.8  & 15.000  &  -30.000   &   1.93  &  306.902  \\
21 & J0152-1637 & 01:52:10.8 & -16:37:52.9  & 3.100   & -27.000  &    0.79  &  101.789  \\
22 & J0206-4028 & 02:06:01.2 & -40:28:04.3  & -10.000  &   75.000 &     0.88  &  315.673  \\
23 & J0218+4232 & 02:18:06.3 & +42:32:17.4  &  C   &       C   &   5.85  &  138.673  \\
24 & J0255-5304 & 02:55:56.2 & -53:04:21.3  & 0.000 &   70.000   &   1.15 &  381.648  \\
25 & J0304+1932 & 03:04:33.1 & +19:32:51.4  & 6.000 &   -37.000  &    0.95  &  168.822  \\
26 & J0323+3944 & 03:23:26.6 & +39:44:52.9  & 16.000  &  -30.000  &    1.49 &   240.177  \\
27 & J0332+5434 & 03:32:59.3 & +54:34:43.5  & 17.000  &   -9.500  &    1.06  &   98.220  \\
28 & J0335+4555 & 03:35:16.6 & +45:55:53.4  &  C   &       C  &    2.08  &   63.143  \\
29 & J0357+5236 & 03:57:44.8 & +52:36:57.7  &  C   &       C   &   4.73  &  242.561  \\
30 & J0358+5413 & 03:58:53.7 & +54:13:13.7  & 9.200 &    8.170  &   1.10  &   64.102  \\
31 & J0406+6138 & 04:06:30.0 & +61:38:40.9  &   C     &     C   &   3.05 &  942.673  \\
32 & J0415+6954 & 04:15:55.6 & +69:54:09.8  &   C    &      C   &   1.57 &    94.151  \\
33 & J0437-4715 & 04:37:15.8 & -47:15:08.8  & 121.310 &   -71.530 &     0.16 &   105.977  \\
34 & J0452-1759 & 04:52:34.0 & -17:59:23.4  & 12.000 &    18.000  &   3.14  &  322.047 \\
35 & J0454+5543 & 04:54:07.7 & +55:43:41.5  & 52.000  &  -17.000   &   0.79  &  204.903  \\
36 & J0502+4654 & 05:02:04.5 & +46:54:06.0  & -8.000  &    8.000  &    1.78 &    95.475 \\
37 & J0525+1115 & 05:25:56.4  &+11:15:19.0  & 30.000   &  -4.000  &    7.68  & 1101.985  \\
38 & J0528+2200 & 05:28:52.3 & +22:00:01    & -20.000   &   7.000  &    2.28  &  229.047  \\
39 & J0534+2200 & 05:34:31.9 & +22:00:52.0  & -14.700   &   2.000  &    2.00   & 140.669  \\
40 & J0538+2817 & 05:38:25.0 & +28:17:09.3  & -23.530   &  52.590  &    1.47 &  401.682 \\
41 & J0543+2329 & 05:43:09.6 & +23:29:05    & 19.000  &   12.000  &    3.54  &  377.151  \\
42 & J0601-0527 & 06:01:58.9&  -05:27:50.5  & 18.000   & -16.000  &    7.54   & 860.899  \\
43 & J0610-2100 & 06:10:13.5 & -21:00:28.0  & 7.000  &   11.000  &    5.64  & 348.635  \\
44 & J0612+3721 & 06:12:48.6 & +37:21:37.3  &  C  &        C  &    1.49   &  49.950  \\
45 & J0613-0200 & 06:13:43.9 & -02:00:47.1  & 2.100  &  -10.500  &    0.48   &  24.174  \\
46 & J0614+2229 & 06:14:17.1 & +22:30:36    & -4.000  &   -3.000  &   4.74   & 112.361  \\
47 & J0621+1002 & 06:21:22.1 & +10:02:38.7  & 3.500  &   -0.300   &   1.88  &   31.310  \\
48 & J0629+2415 & 06:29:05.7 & +24:15:43.3  & -7.000  &    2.000  &    4.67  &  161.184  \\
49 & J0630-2834 & 06:30:49.4 & -28:34:43.1  & -44.600  &   20.000  &    2.15   & 498.228 \\
50 & J0633+1746 & 06:33:54.1 & +17:46:12.9  & 138.000  &   97.000 &   0.16 &  124.953  \\
51 & J0653+8051 & 06:53:15.0 & +80:52:00.2  & 19.000  &   -1.000   &   3.37  &  303.984  \\
52 & J0659+1414 & 06:59:48.1 & +14:14:21.5  & 44.070  &   -2.400   &   0.29  &   60.300 \\
53 & J0700+6418 & 07:00:37.8 & +64:18:11.2  &   C   &       C   &   0.48  &   22.757 \\
54 & J0711-6830 & 07:11:54.2 & -68:30:47.4  & -11.000   &  19.000  &    1.04  &  108.249  \\
55 & J0737-3039A & 07:37:51.2 & -30:39:40.7 & -3.300  &    2.600  &    0.33  &    6.639  \\
56 & J0737-3039B & 07:37:51.2 & -30:39:40.7 & -3.300  &    2.600  &    0.33  &    6.639  \\
57 & J0738-4042  & 07:38:32.3 & -40:42:40.9 & -14.000  &   13.000  &   11.03  &  999.054  \\
58 & J0742-2822 & 07:42:49.0 & -28:22:43.7  & -29.000    &  4.000 &     1.89  &  262.313  \\
59 & J0751+1807 & 07:51:09.1 & +18:07:38.6  &   C   &       C  &    0.62   &  17.809  \\
60 & J0754+3231 & 07:54:40.6  &+32:31:56.2  & -4.000  &    7.000   &   3.92  &  149.834  \\
61 & J0758-1528 & 07:58:29.0 & -15:28:08.7  & 1.000   &   4.000  &    3.72  &   72.717  \\
62 & J0814+7429 & 08:14:59.5 & +74:29:05.7  & 24.020   & -44.000   &   0.43  &  102.883 \\
63 & J0820-1350 & 08:20:26.3 & -13:50:55.4  & 9.000  &  -47.000  &    2.45  &  555.841 \\
64 & J0823+0159 & 08:23:09.7  &+01:59:12.4  & 5.000  &   -1.000  &    1.44  &   34.811  \\
65 & J0826+2637 & 08:26:51.3  &+26:37:23.7  & 61.000   & -90.000 &     0.36  &  184.091  \\
66 & J0835-4510 & 08:35:20.6 & -45:10:34.8  & -49.680  &   29.900  &   0.29   &  78.542  \\
67 & J0837+0610 & 08:37:05.6 & +06:10:14.5  & 2.000  &   51.000  &    0.72  &  174.222  \\
68 & J0837-4135 & 08:37:21.1 & -41:35:14.3  & -2.300  &  -18.000  &    4.24   & 364.772 \\
69 & J0846-3533 & 08:46:06.0 & -35:33:40.7  & 93.000    &-15.000   &   1.43 &   638.649  \\
70 & J0908-1739 & 09:08:38.1 & -17:39:37.6  & 27.000  &  -40.000 &     0.63  &  144.143  \\
71 & J0922+0638 & 09:22:14.0 & +06:38:23.3  & 18.800  &   86.400 &     1.20  &  505.062  \\
72 & J0943+1631 & 09:43:30.1 & +16:31:37    & 23.000   &   9.000  &    1.76  &  206.084  \\
73 & J0944-1354 & 09:44:28.9 & -13:54:41.6  & -1.000  &  -22.000   &   0.69   &  72.042  \\
74 & J0946+0951 & 09:46:07.6 & +09:51:55    & -38.000   & -21.000   &   0.98  &  201.720  \\
75 & J0953+0755 & 09:53:09.3  &+07:55:35.7  & -2.090   &  29.460   &   0.26  &  36.654  \\
76 & J1012+5307 & 10:12:33.4 & +53:07:02.5  & 2.400  &  -25.200   &   0.52   &  62.407  \\
77 & J1024-0719 & 10:24:38.6 & -07:19:19.1  & -34.900  &  -47.000  &    0.53  &  146.072  \\
78 & J1041-1942 & 10:41:36.1 & -19:42:13.6  & -1.000  &   14.000  &    3.18 &   211.606  \\
79 & J1045-4509 & 10:45:50.1 &-45:09:54.1   & -7.000   &   8.000  &    3.24  &  163.287 \\
80 & J1115+5030 & 11:15:38.4 & +50:30:12.2  & 22.000  &  -51.000   &   0.54 &   142.196  \\
81 & J1116-4122 & 11:16:43.0 & -41:22:43.9  & -1.000   &   7.000   &   2.68  &   89.843  \\
82 & J1136+1551 & 11:36:03.2 & +15:51:04.4  & -74.000 &   368.100  &    0.36  &  635.733  \\
83 & J1239+2453 & 12:39:40.4 & +24:53:49.2  & -104.500  &   49.400   &   0.86  &  472.410  \\
84 & J1300+1240 & 13:00:03.0 & +12:40:56.7  & 46.400  &  -82.200 &     0.77  &  344.234  \\
85 & J1321+8323 & 13:21:46.1 & +83:23:38.9  & -53.000   &  13.000  &    0.77&    199.214  \\
86 & J1328-4357 & 13:28:06.4 & -43:57:44.1  & 3.000  &   54.000  &    2.29  &  587.172  \\
87 & J1430-6623 & 14:30:40.8 & -66:23:05.0  & -31.000  &  -21.000  &    1.80  &  319.531  \\
88 & J1453-6413 & 14:53:32.7 & -64:13:15.5  & -16.000  &  -21.300   &   1.84 &   232.391  \\
89 & J1455-3330 & 14:55:47.9& -33:30:46.3   & 5.000  &  24.000   &   0.74  &   86.007  \\
90 & J1456-6843 & 14:56:00.1 & -68:43:39.2  & -39.500  &  -12.300  &    0.45  &   89.153  \\
91 & J1509+5531 & 15:09:25.6 & +55:31:32.3  &-73.606  &  -62.622   &   2.41  & 1104.013  \\
92 & J1518+4904 & 15:18:16.7 & +49:04:34.2  &  C    &      C  &   0.70  &   30.914  \\
93 & J1537+1155 & 15:37:09.9&  +11:55:55.5  & 1.340 &   -25.050 &     0.90  &  107.038  \\
94 & J1543-0620 & 15:43:30.1 & -06:20:45.2  & -17.000   &  -4.000 &     1.15  &   95.217  \\
95 & J1543+0929 & 15:43:38.8 & +09:29:16.5  & -7.300 &    -4.000 &     2.46  &   97.082  \\
96 & J1559-4438 & 15:59:41.5 & -44:38:46.1  &  1.000  &   14.000  &    1.58  &  105.137  \\
97 & J1600-3053 & 16:00:51.9 & -30:53:49.3  & -1.020 &   -6.700  &    2.67  &   85.788  \\
98 & J1603-7202 & 16:03:35.6 & -72:02:32.6  & -2.800 &    -6.000  &    1.64  &   51.481  \\
99 & J1604-4909 & 16:04:22.9 & -49:09:58.3  & -30.000  &   -1.000  &    3.59  &  510.886  \\
100 & J1607-0032 & 16:07:12.1 & +00:32:40.8 & -1.000  &   -7.000  &    0.59  &   19.779  \\
101 & J1623-2631 & 16:23:38.2 & -26:31:53.7 & -13.400  &  -25.000 &     2.20 &   295.848  \\
102 & J1640+2224 & 16:40:16.7 & +22:24:08.9 & 1.660  &  -11.300 &     1.19 &    64.436  \\
103 & J1643-1224 & 16:43:38.1 & -12:24:58.7 & 3.000  &   -8.000  &    4.86  &  196.863  \\
104 & J1645-0317 & 16:45:02.0 & -03:17:58.3 & -3.700  &   30.000  &    2.91  &  417.022  \\
105 & J1709-1640 & 17:09:26.4 & -16:40:57.7 & 3.000  &    0.000  &    1.27 &    18.063  \\
106 & J1709+2313 & 17:09:05.7 & +23:13:27.8 & -3.200  &   -9.700  &    1.83  &   88.618 \\
107 & J1713+0747 & 17:13:49.5 & +07:47:37.5 & 4.917  &   -3.933  &    1.12  &  33.541  \\
108 & J1720-0212 & 17:20:57.2 & -02:12:23.9 & -1.000 &   -26.000 &     5.41  &  667.357  \\
109 & J1722-3207 & 17:22:02.9 & -32:07:45.3  & -1.000 &   -40.000 &     3.18 &  603.239  \\
110 & J1735-0724 & 17:35:04.9 & -07:24:52.4  & -2.400  &   28.000 &     4.32  &  575.570 \\
111 & J1738+0333 & 17:38:53.9 & +03:33:10.8  & 5.600 &     4.000   &   1.97  &   64.275 \\
112 & J1740+1311 & 17:40:07.3 & +13:11:56.6  & -22.000  &  -20.000   &   4.77  &  672.375  \\
113 & J1741-3927 & 17:41:18.0 & -39:27:38.0  & 20.000 &    -6.000  &    4.75 &   470.223 \\
114 & J1744-1134 & 17:44:29.4 & -11:34:54.6  & 19.600  &   -7.000  &    0.48  &   46.986  \\
115 & J1745-3040 & 17:45:56.3 & -30:40:23.5  & 6.000 &     4.000  &    2.08 &    71.110  \\
116 & J1752-2806 & 17:52:58.6 & -28:06:37.3  & -4.000 &    -5.000  &    1.53 &    46.446  \\
117 & J1801-2451 & 18:00:59.8 & -24:51:53    & 2.000   &  -3.000 &     4.61  &   78.802 \\
118 & J1801-1417 & 18:01:51.0 & -14:17:34.5  & -8.000  &  -23.000  &   1.80  &  207.810  \\
119 & J1803-2137 & 18:03:51.4 & -21:37:07.3  & 11.600   &  14.800  &    3.94  &  351.252  \\
120 & J1807-0847 & 18:07:38.0 & -08:47:43.2  & -5.000   &   1.000  &    3.60  &   87.027 \\
121 & J1820-0427 & 18:20:52.6 & -04:27:38.1  &   C   &       C &     2.45 &   164.266  \\
122 & J1823+0550 & 18:23:30.9 & +05:50:24.3  & 5.000  &   -2.000  &    3.01  &   76.848  \\
123 & J1824-1945 & 18:24:00.4 & -19:45:51.7  & -12.000  & -100.000  &    5.20 &  2482.989  \\
124 & J1824-2452 & 18:24:32.0 & -24:52:11.1  & -0.900  &   -4.600  &    4.90 &   108.887  \\
125 & J1825-0935 & 18:25:30.5 & -09:35:22.1  & -13.000  &   -9.000 &     1.00  &   74.961  \\
126 & J1829-1751 & 18:29:43.1 & -17:51:03.9  & 22.000  & -150.000 &    5.49 &  3945.954  \\
127 & J1832-0827 & 18:32:37.0 & -08:27:03.6  & -4.000  &   20.000 &    4.75 &   459.311  \\
128 & J1833-0827 & 18:33:40.3 & -08:27:31.2  &  C  &        C  &    5.67  &  901.625  \\
129 & J1835-1106 & 18:35:18.2 & -11:06:15.1  & 27.000   &  56.000 &     3.08  &  907.804  \\
130 & J1836-0436 & 18:36:51.7 & -04:36:37.6  &   C &         C   &  4.62 &   265.780   \\
131 & J1836-1008 & 18:36:53.9 & -10:08:08.3  & 18.000   &  12.000  &    5.39 &   552.813  \\
132 & J1840+5640 & 18:40:44.6 & +56:40:55.4  & -30.000   & -21.000  &    1.70  &  295.141  \\
133 & J1841-0425 & 18:41:05.6 & -04:25:19.6  &   C   &       C   &   5.17 &   285.843  \\
134 & J1844+1454 & 18:44:54.8 & +14:54:14.1  & -9.000   &  45.000 &    2.23  &  485.178  \\
135 & J1850+1335 & 18:50:35.4 & +13:35:58.3  &   C  &        C  &    3.14  &  240.040  \\
136 & J1857+0943 & 18:57:36.3 & +09:43:17.2  & -2.940   &  -5.410 &    0.91  &   26.537  \\
137 & J1900-2600 & 19:00:47.5 & -26:00:43.8  & -19.900  &  -47.300   &   2.00  &  486.568 \\
138 & J1902+0615 & 19:02:50.2 & +06:16:33.4  &   C       &   C    &  9.63 &   368.086  \\
139 & J1905-0056 & 19:05:27.7 & -00:56:40.9  &   C   &       C  &    6.91  &  296.655  \\
140 & J1906+0641 & 19:06:35.2 & +06:41:02.9  &   C   &       C  &    9.23 &  269.607  \\
141 & J1907+4002 & 19:07:34.6 & +40:02:05.7  & 11.000  &   11.000  &    1.76 &   129.804  \\
142 & J1909+1102 & 19:09:48.6 & +11:02:03.3  &   C   &       C  &    4.31 &   192.770  \\
143 & J1909-3744 & 19:09:47.4 & -37:44:14.3  & -9.490   & -36.060 &     1.14 &   200.885  \\
144 & J1910-5959A & 19:11:42.7 & -59:58:26.9 & -3.300  &   -3.600  &    4.00  &   92.613  \\
145 & J1910-5959C & 19:11:05.5&  -60:00:59.7 & -4.100  &   -4.600   &   4.00  &  116.855  \\
146 & J1911-1114 & 19:11:49.2 & -11:14:22.3  & -6.000  &  -23.000  &    1.59 &   179.179  \\
147 & J1913+1400 & 19:13:24.3 & +14:00:52.7  &    C   &       C  &    5.07  &  326.934  \\
148 & J1913-0440 & 19:13:54.1 & -04:40:47.6  & 7.000   &  -5.000  &    3.22  &  131.322  \\
149 & J1915+1009 & 19:15:29.9 & +10:09:43.7  &    C   &       C   &   5.32 &   177.003  \\
150 & J1915+1606 & 19:15:27.9 & +16:06:27.4  & -2.560  &    0.490  &   7.13  &   88.107  \\
151 & J1916+0951 & 19:16:32.3 & +09:51:25.9  &    C   &       C  &    2.88 &   123.257  \\
152 & J1917+1353 & 19:17:39.7 & +13:53:56.9  & 0.000  &   -6.000 &    4.07  &  115.774  \\
153 & J1919+0021 & 19:19:50.6 & +00:21:39.8  & -2.000   &  -1.000 &    3.32  &   35.196  \\
154 & J1921+2153 & 19:21:44.8 & +21:53:02.2  & 17.000 &   32.000  &    0.66 &   113.382  \\
155 & J1926+1648 & 19:26:45.3 & +16:48:32.7  &    C  &        C  &    7.74 &   597.353  \\
156 & J1932+1059 & 19:32:13.9 & +10:59:32.4  & 94.090   &  42.990 &     0.36  &  177.051  \\
157 & J1935+1616 & 19:35:47.8 & +16:16:40.2  & -1.000  & -13.000   &   7.93 &  490.190  \\
158 & J1937+2544 & 19:37:01.2 & +25:44:13.6  &    C   &       C  &    2.76 &   232.973  \\
159 & J1939+2134 & 19:39:38.5 & +21:34:59.1  & -0.460  &   -0.660  &    3.60  &   13.731  \\
160 & J1941-2602 & 19:41:00.4 & -26:02:05.7  & 12.000  &  -10.000  &    4.74  &  351.027 \\
161 & J1944+0907 & 19:44:09.3 & +09:07:23.2  & 12.000  &  -18.000 &     1.28  &  131.280  \\
162 & J1946-2913 & 19:46:51.7 & -29:13:47.1  & 19.000 &  -33.000  &    4.31  &  778.087 \\
163 & J1946+1805 & 19:46:53.0 & +18:05:41.2  & 1.000  &   -9.000  &    0.85  &   36.492 \\
164 & J1948+3540 & 19:48:25.0 & +35:40:11.0  & -12.600  &    0.700  &    7.87  &  470.849  \\
165 & J1952+3252 & 19:52:58.2 & +32:52:40.5  & -24.000  &   -8.000  &    2.50 &   299.845  \\
166 & J1954+2923 & 19:54:22.5 &+29:23:17.2   & 25.000   & -36.000    &  0.42 &    87.273  \\
167 & J1955+2908 & 19:55:27.8 & +29:08:43.5  & -1.000   &  -3.700   &   5.39  &   97.941 \\
168 & J1955+5059 & 19:55:18.7 & +50:59:55.2  & -23.000 &   54.000  &    1.80 &   500.880 \\
169 & J1959+2048 & 19:59:36.7 & +20:48:15.1  & -16.000  &  -25.800  &    1.53  &  220.211  \\
170 & J2002+4050 & 20:02:44.0 &  +40:50:53.9 &   C   &       C  &    8.30  &  524.995  \\
171 & J2004+3137 & 20:04:52.2 & +31:37:10.0  &   C   &       C   &   8.94 &   473.870  \\
172 & J2013+3845 & 20:13:10.3 & +38:45:43.3  & -32.100   & -25.000 &    13.07 &  2521.130  \\
173 & J2018+2839 & 20:18:03.8 & +28:39:54.2  & -2.600 &    -6.200 &     0.97  &   30.945  \\
174 & J2019+2425 & 20:19:31.9 & +24:25:15.3  & -9.410  &  -20.600 &     0.91  &   97.708  \\
175 & J2022+2854 & 20:22:37.0 & +28:54:23.1  & -4.400  &  -23.600 &     2.70 &   307.605  \\
176 & J2022+5154 & 20:22:49.8 & +51:54:50.2  & -5.230 &    11.500  &    2.00 &  119.788  \\
177 & J2023+5037 & 20:23:41.9 & +50:37:34.8  &   C  &        C    &  1.80  &  187.936  \\
178 & J2046-0421 & 20:46:00.1 & -04:21:26.0  & 9.000  &   -7.000  &    3.83 &   207.032  \\
179 & J2046+1540 & 20:46:39.3 & +15:40:33.6  & -13.000  &    3.000&      2.56 &   161.926  \\
180 & J2048-1616 & 20:48:35.4 & -16:16:43.0  & 117.000  &   -5.000 &     0.64 &   355.328  \\
181 & J2051-0827 & 20:51:07.5 & -08:27:37.7  & 5.300  &    0.300  &    1.28  &   32.214  \\
182 & J2055+2209 & 20:55:39.1 & +22:09:27.2  &   C &        C  &   2.15  &   52.413  \\
183 & J2055+3630 & 20:55:31.3 & +36:30:21.4  & -3.000   &   3.000 &    5.56  &  111.835  \\
184 & J2108+4441 & 21:08:20.4 & +44:41:48.8  & 3.500   &   1.400 &     5.28  &   94.362  \\
185 & J2113+2754 & 21:13:04.3 & +27:54:02.2  & -23.000   & -54.000 &    1.39 &  386.791  \\
186 & J2113+4644 & 21:13:24.3 & +46:44:08.7  &    C   &       C  &    4.99 &   308.455  \\
187 & J2116+1414 & 21:16:13.7 & +14:14:21.0  & 8.000  &  -11.000 &    4.43  &  285.665  \\
188 & J2124-3358 & 21:24:43.8 & -33:58:44.6  & -14.400   & -50.000 &     0.25 &    61.670  \\
189 & J2129+1210A & 21:29:58.2 & +12:10:01.2 & -0.260  &   -4.400 &     8.90  &  185.980  \\
190 & J2129+1210B & 21:29:58.6 & +12:10:00.3 & 1.700  &   -1.900 &     8.90  &  107.576  \\
191 & J2129+1210C & 21:30:01.2 & +12:10:38.2 & -1.300  &   -3.300  &   10.00  &  168.154  \\
192 & J2129-5721 & 21:29:22.7 & -57:21:14.1  & 7.000   &  -4.000  &    2.55   &  97.468  \\
193 & J2145-0750 & 21:45:50.4 & -07:50:18.3  &  C   &      C  &   0.50  &  33.461  \\
194 & J2149+6329 & 21:49:58.5 & +63:29:43.5  & 14.000  &   10.000 &    13.65 &  1113.385  \\
195 & J2150+5247 & 21:50:37.7 & +52:47:49.6  &  C   &       C  &    5.67  &  231.241  \\
196 & J2157+4017 & 21:57:01.8 & +40:17:45.8  & 17.800 &     2.800 &     5.58 &   476.682  \\
197 & J2219+4754 & 22:19:48.1 & +47:54:53.9  & -12.000  &  -30.000  &    2.45  &  375.304  \\
198 & J2225+6535 & 22:25:52.4 & +65:35:34.0  & 144.000  &  112.000  &    2.00 &  1729.770  \\
199 & J2229+6205 & 22:29:41.8 & +62:05:36.0  &   C  &        C  &    5.70 &   270.235  \\
200 & J2229+2643 & 22:29:50.8 & +26:43:57.7  & 1.000  &  -17.000  &    1.43  &  115.452  \\
201 & J2235+1506 & 22:35:43.7 & +15:06:49.0  & 15.000 &    10.000  &    1.15  &   98.289 \\
202 & J2257+5909 & 22:57:57.7 & +59:09:14.8  &   C   &       C  &    6.40  &  540.227  \\
203 & J2257+5909 & 22:57:57.7 & +59:09:14.8  &   C  &        C   &   6.40  &  540.227  \\
204 & J2305+3100 & 23:05:58.3 & +31:00:01.7  & 2.000  &  -20.000   &   3.92  &  373.546 \\
205 & J2308+5547 & 23:08:13.8 & +55:47:36.0  & -15.000  &    0.000  &    2.42  &  172.097  \\
206 & J2313+4253 & 23:13:08.5 & +42:53:12.9  &   C  &       C    &  0.95  &   99.296  \\
207 & J2317+1439 & 23:17:09.2 & +14:39:31.2  & -1.700  &    7.400   &   1.89 &    68.034  \\
208 & J2321+6024 & 23:21:55.2& +60:24:30.7   & -17.000 &    -7.000 &     3.21 &   279.789  \\
209 & J2322+2057 & 23:22:22.3 & +20:57:02.9  & -17.000  &  -18.000   &   0.78 &    91.557  \\
210 & J2326+6113 & 23:26:58.6 & +61:13:36.4  &   C  &        C  &    4.82  &  433.576  \\
211 & J2330-2005 & 23:30:26.8 & -20:05:29.6  & 74.700  &    5.000   &   0.49 &   173.922  \\
212 & J2337+6151 & 23:37:05.7 & +61:51:01.6  &   C    &      C    &  2.47 &   124.481  \\
213 & J2354+6155 & 23:54:04.7 & +61:55:46.7  & 22.000  &    6.000 &  3.31  &  357.846  \\
\hline \hline
\end{longtable*}

{ From the analysis of this Table one can realize that most of the SNR that were 
obtained gather about an $S/N \sim {\cal{O}}(1)$. However, an SNR of the 
order of unity is not enough for detection. The statistics which is the output 
of the {\sl matched filter} must stand over the GW detector noise. Thus, if 
one assumes Gaussian noise in the detector and a Gaussian distributed statistics,
then an SNR of something like 5, {\sl what means a chance of one in a million}, 
is required to state that {\sl it is signal and not noise}. Notwithstanding, 
RAPs current kinematics suggest that the dynamical situation could have been 
dramatically different during the short rise fling that launches each pulsar 
to drift across the Galaxy. In virtue of the potentiality of this phenomenon
for the emission of powerful gravitational radiation signals, in the next 
Section we analyze the pulsar kick at birth within the framework of a theory 
that appears more appropriate to describe that process, and reestimate the 
characteristics of the GW released. }


\section{Gravitational waves from pulsar acceleration}

{ Observations of extragalactic astrophysical sources suggest that blobs of matter 
are ejected with ultrarelativistic speeds (Lorentz factor $\gamma >> 1$) in various 
powerful phenomena involving quasars, active galactic nuclei, radio-galaxies, supernova 
explosions, and microquasars. For these sources the emission of gravitational waves 
when such an ultrarelativistic blob is ejected from the core of its host source 
have been studied by Segalis and Ori\cite{ori2001} (see also \cite{sago04} for 
the analysis of similar phenomena in gamma-ray bursts). 

Supernova explosions are known to be the natural place of birth of neutron stars 
and pulsars, e. g. {\sl The Crab Nebula}. In processes such these, the nascent 
pulsar is kicked out from the supernova remnant with speed that is not so large
as compared to that for the blobs in AGN or quasars, where a significant fraction 
of the speed of light is typical for the ejecta. Nonetheless, the velocity at birth
can be considered large if one bears in mind that the typical mass of neutron stars 
is about one solar mass\cite{chakra-thorsett}. Although a galactic RAP cannot be 
considered an ultrarelativistic source, the pulsar rise fling during the supernova 
can kick it out with very high speed. That is why in the discussion below we analyze 
the gravitational radiation emitted during the kick at birth of RAPs within the 
framework of Ref.\cite{ori2001}. Besides, it is also of worth to keep in mind that 
such an approach can be applied to the early RAPs dynamics because in this case the 
limit 
$\gamma \sim 1$ applies. This limit corresponds to the GW strain $h$ provided by the 
well-known general relativistic quadrupole formula for the same dynamical situation 
\cite{braginsky-thorne83}. But bear in mind that this does not describe properly
the GW angular distribution.}


{

The pulsar is envisioned hereafter as a ``particle"  of mass $M_\star$ moving 
along the worldline $r^\lambda(\tau)$ (with $\tau$ the proper time)  and having 
an energy-stress tensor

\begin{equation}
T_{\mu \nu}(x) = M_\star \int V_{\mu} V_{\nu} \delta^{(4)}[ x -
r(\tau)]~d\tau \; ,
\end{equation}

where $V^\alpha = dr^\alpha/d\tau$ is the particle 4-vector velocity.
(Upper[sub])scripts are raised[lowered] with the Minkowski metric
$\eta_{\mu \nu}$).

Since we are using the linearized Einstein's equations to describe
the emission of GW during the early acceleration phase of any
pulsar, then the resulting metric perturbation produced by the RAP
can straightforwardly be evaluated at the retarded time, which
corresponds to the intersection time of $r^\alpha(\tau)$ with the
observer's past light-cone, and then be transformed to the Lorentz
gauge where it reads:  ${h}_{\mu \nu} = \bar{h}_{\mu\nu} -
\frac{1}{2} \eta_{\mu \nu}\bar{h}_{\alpha}^\alpha$. This strain
can equivalently be written as

\begin{equation}
{h}_{\mu \nu} = \frac{4~M_\star}{- V_\lambda \cdot [ x - r(\tau)]^\lambda }
\left(V_\mu(\tau) V_\nu(\tau) + \frac{1}{2} \eta_{\mu \nu} \right) \; .
\label{eff-strain}
\end{equation}

Notice that is this factor $-V_\lambda \cdot[ x - r(\tau)]^\lambda
$; which depends on the velocity in the denominator of
Eq.(\ref{eff-strain}), that is responsible for the nonvanishing
amplitude (GW ``memory'' \cite{braginsky-thorne83}) of the GW
signal produced by the launching of the pulsar into its present
trajectory.

}

Eq.(\ref{eff-strain}) must be finally rewritten in the
tranverse-traceless ($TT$) gauge, i. e. ${h}_{\mu \nu} \longrightarrow
{h}^{TT}_{\mu  \nu}$; which is the best suited to discuss the GW
detector's response to that  signal. This procedure leads to the result
presented in Eq.(\ref{hmax}).




{ A detailed analysis (see Refs.\cite{ori2001,braginsky-thorne83}) shows 
that the maximum GW strain in the detector is obtained for a wavevector, 
$\vec{n}$, orthogonal to the detector's arm, in the $\vec{v}$ - $\vec{n}$ 
plane. Here $\vec{n}$ is the unit spatial direction vector from the (retarded) 
pulsar position to the observer, and $\vec{v}$ the unit spatial direction 
vector defining the 3-D pulsar velocity $\vec{V}$. In this case, the GW 
amplitude generated by the pulsar kick becomes (c.g.s. units recovered)

\begin{equation}
h_{\rm max}(t) = \left[\frac{G}{c^2}\right] \int_0^{\theta_V} 2 
\frac{\gamma(t) M_\star \beta^2(t)}{D_\star} \left(\frac{\sin^3\theta 
\sin2\Delta\phi}{\Delta\Omega (1 - \beta \cos\theta)} \right) 
~d\theta\; , 
\label{hmax}
\end{equation}

where $\theta$ is the angle between $\vec{v}$ and $\vec{n}$, that is 
the angle between the unit spatial direction vector associated to velocity 
of the pulsar, $\vec{V}$, and the unit spatial direction vector linking the
source point to the observer location, i. e. $\vec{v} \cdot \vec{n} = 
\cos\theta$). $\theta_V$ is the particular (e. g. {\sl Vela Pulsar}) viewing 
angle that a RAP velocity makes with the direction $\vec{n}$ to the 
observer. $\Delta \theta$ is the angular diameter of the pulsar as 
seen by the observer. (For galactic distances it can be assumed nearly 
constant for each of the RAPs in the sample under analysis). Thus 
$\Delta \Omega \simeq \pi (\Delta \theta)^2$ is the solid angle over 
which the RAP is viewed. Besides, $\beta = |\vec{V}|/c$, being $|\vec{V}|$ 
the pulsar 3-D velocity $V^{\rm fin}_\star (t)$ defined above, and $\gamma$ 
the Lorentz factor; which can be taken here as $\sim 1$ because $V^{\rm 
fin}_\star(t) \ll c$. It is also defined $\Delta \phi = \cos^{-1}\left(
\frac{\cos \Delta\theta - \cos\theta_V \cos\theta } {\sin \theta_V \sin 
\theta} \right)$. Finally, the distance to the pulsar is $D_\star$. 

This formula was obtained in the form presented above first by Sago et 
al.\cite{sago04} in discussing the GW emission from gamma-ray bursts (GRBs). 
It is essentially equivalent to the formula obtained by Segalis and Ori
\cite{ori2001} in an earlier paper in which they discussed the GW 
emission by ultrarelativistic sources. The main different in between 
is related to the detailed description of the angular distribution of
the gravitational radiation emitted in each of the cases that those 
authors analyzed. The formula indicates that the GW strain depends 
on the Lorentz factor, pulsar speed and distance to the observer, and 
also on the strong beaming effect in the case of GRBs, as indicated by 
the factors $(1 - \cos\theta)$, and $\Delta\Omega$, appearing in the 
denominator of the equation. These effects are no so much important
in the case of Galactic RAPs, as is evident from the pulsar speed
and angular size.
}

This result, shown in Fig.\ref{GW-PSR-GRB-SIGNAL}, states  that the GW 
space-time perturbation is not strongly beamed in the forward direction 
$\vec{n}$, as opposed to the case of the electromagnetic radiation in
gamma-ray bursts (also shown in the same Figure). Instead, the 
metric perturbation at the ultra-relativistic limit (not applicable to 
RAPs) has a directional dependence scaling as  $1 + \cos  \theta$. In
such a case, because of the strong beaming effect, as in gamma-ray
bursts; for instance, the electromagnetic radiation emitted by the
source over the same time interval is visible only inside the very
small solid angle $(\theta \sim \gamma^{-1})^2$, whereas the GW signal
is observable within a wider solid angle; almost $2 \pi$ radians (see
Fig.\ref{GW-PSR-GRB-SIGNAL}). Besides, the observed GW frequency is 
Doppler blueshifted in the forward direction, and  therefore the 
energy flux carried by the GWs is beamed in  the forward direction, 
too. Quite noticeable, in the case of RAPs the GW signal will have its 
maximum for viewing angles $\theta \sim \pi/2$, that is; for pulsar 
motions purely in the plane of the sky, as shown in Fig.
\ref{GW-PSR-GRB-SIGNAL}. This key physical property of the emission 
process allows us to neglect the radial component of the pulsar 
velocity because it does not contribute to the effective GW amplitude 
of the signal being emitted by the RAP, or in other words, to the signal 
to be detected on Earth.




\begin{longtable}{llll}
\caption{\label{gw-characteristics} Characteristic amplitude and frequency, 
$h_c$, $f_c$(Hz), and $ S/N = h_c(f_c)/h_{\textrm rms}(f_c)$.}\\
\hline \hline
Jname & f$_c$ &  h$_c$ & S/N \\
 { }  & [Hz]  &  { }   &  {  } \\
\hline 
\hline
\endfirsthead
\caption{results of full calculation of $h_c$, $f_c$ and $S/N$.}\\
\hline \hline
Jname & f$_c$ &  h$_c$ & S/N \\
 { }  & [Hz]  &  { }   &  {  } \\
\hline 
\hline
\endhead
\text{J0014+4746} & 37.1239 & 7.99071$ \times 10^{-24}$ & 0.465967 \\
 \text{J0024-7204C} & 39.0641 & 1.81774$ \times 10^{-24}$ & 0.114416 \\
 \text{J0024-7204D} & 43.0236 & 1.13491$ \times 10^{-24}$ & 0.0810649 \\
 \text{J0024-7204E} & 38.1603 & 2.03054$ \times 10^{-24}$ & 0.123401 \\
 \text{J0024-7204F} & 40.5101 & 1.50882$ \times 10^{-24}$ & 0.100293 \\
 \text{J0024-7204G} & 40.1852 & 1.57348$ \times 10^{-24}$ & 0.103335 \\
 \text{J0024-7204H} & 39.0137 & 1.82934$ \times 10^{-24}$ & 0.114923 \\
 \text{J0024-7204I} & 38.98 & 1.83713$ \times 10^{-24}$ & 0.115263 \\
 \text{J0024-7204J} & 38.8167 & 1.87205$ \times 10^{-24}$ & 0.116717 \\
 \text{J0024-7204N} & 37.9712 & 2.07818$ \times 10^{-24}$ & 0.125358 \\
 \text{J0024-7204O} & 40.5101 & 1.50882$ \times 10^{-24}$ & 0.100293 \\
 \text{J0024-7204U} & 38.1007 & 2.04539$ \times 10^{-24}$ & 0.124012 \\
 \text{J0030+0451} & 79.288 & 2.32394$ \times 10^{-25}$ & 0.0165996 \\
 \text{J0034-0534} & 38.6853 & 9.31883$ \times 10^{-24}$ & 0.578055 \\
 \text{J0040+5716} & 37.6624 & 2.31387$ \times 10^{-24}$ & 0.137876 \\
 \text{J0117+5914} & 36.6742 & 5.64571$ \times 10^{-24}$ & 0.323257 \\
 \text{J0134-2937} & 36.5873 & 6.90089$ \times 10^{-24}$ & 0.393722 \\
 \text{J0139+5814} & 38.4259 & 5.93842$ \times 10^{-24}$ & 0.364667 \\
 \text{J0147+5922} & 43.9134 & 2.67217$ \times 10^{-24}$ & 0.19087 \\
 \text{J0151-0635} & 38.4397 & 9.25843$ \times 10^{-24}$ & 0.568847 \\
 \text{J0152-1637} & 43.0987 & 6.85662$ \times 10^{-24}$ & 0.489759 \\
 \text{J0206-4028} & 38.5556 & 2.08228$ \times 10^{-23}$ & 1.28516 \\
 \text{J0218+4232} & 39.2696 & 1.45314$ \times 10^{-24}$ & 0.0921892 \\
 \text{J0255-5304} & 39.0962 & 1.86966$ \times 10^{-23}$ & 1.1783 \\
 \text{J0304+1932} & 37.2766 & 1.14724$ \times 10^{-23}$ & 0.673126 \\
 \text{J0323+3944} & 37.1671 & 9.82703$ \times 10^{-24}$ & 0.57405 \\
 \text{J0332+5434} & 43.6786 & 4.89642$ \times 10^{-24}$ & 0.349744 \\
 \text{J0335+4555} & 53.156 & 1.38012$ \times 10^{-24}$ & 0.09858 \\
 \text{J0357+5236} & 37.143 & 3.12732$ \times 10^{-24}$ & 0.182506 \\
 \text{J0358+5413} & 52.7994 & 2.66806$ \times 10^{-24}$ & 0.190576 \\
 \text{J0406+6138} & 41.529 & 1.51989$ \times 10^{-23}$ & 1.04865 \\
 \text{J0415+6954} & 44.4139 & 3.13942$ \times 10^{-24}$ & 0.224244 \\
 \text{J0437-4715} & 42.4872 & 3.55097$ \times 10^{-23}$ & 2.53527 \\
 \text{J0452-1759} & 38.6709 & 5.93204$ \times 10^{-24}$ & 0.367763 \\
 \text{J0454+5543} & 36.4967 & 1.65967$ \times 10^{-23}$ & 0.943389 \\
 \text{J0502+4654} & 44.1655 & 2.817$ \times 10^{-24}$ & 0.201214 \\
 \text{J0525+1115} & 40.4211 & 7.49499$ \times 10^{-24}$ & 0.496559 \\
 \text{J0528+2200} & 36.9714 & 6.20109$ \times 10^{-24}$ & 0.359381 \\
 \text{J0534+2200} & 39.1178 & 4.33307$ \times 10^{-24}$ & 0.273303 \\
 \text{J0538+2817} & 39.3889 & 1.50976$ \times 10^{-23}$ & 0.962179 \\
 \text{J0543+2329} & 39.151 & 6.00026$ \times 10^{-24}$ & 0.378942 \\
 \text{J0601-0527} & 42.5181 & 5.41622$ \times 10^{-24}$ & 0.386873 \\
 \text{J0610-2100} & 38.8927 & 3.54204$ \times 10^{-24}$ & 0.221485 \\
 \text{J0612+3721} & 58.5628 & 1.34323$ \times 10^{-24}$ & 0.0959447 \\
 \text{J0613-0200} & 71.1509 & 1.16838$ \times 10^{-24}$ & 0.0834554 \\
 \text{J0614+2229} & 41.9176 & 1.30377$ \times 10^{-24}$ & 0.091219 \\
 \text{J0621+1002} & 67.466 & 4.79347$ \times 10^{-25}$ & 0.0342391 \\
 \text{J0629+2415} & 37.6964 & 2.21027$ \times 10^{-24}$ & 0.131882 \\
 \text{J0630-2834} & 40.8252 & 1.17533$ \times 10^{-23}$ & 0.790392 \\
 \text{J0633+1746} & 40.4001 & 4.59112$ \times 10^{-23}$ & 3.03935 \\
 \text{J0653+8051} & 38.4132 & 5.25441$ \times 10^{-24}$ & 0.322503 \\
 \text{J0659+1414} & 54.243 & 9.24396$ \times 10^{-24}$ & 0.660283 \\
 \text{J0700+6418} & 71.8931 & 1.04344$ \times 10^{-24}$ & 0.0745312 \\
 \text{J0711-6830} & 42.4243 & 5.59481$ \times 10^{-24}$ & 0.398562 \\
 \text{J0737-3039A} & 80.003 & 1.37444$ \times 10^{-25} $& 0.00981745 \\
 \text{J0737-3039B} & 80.003 & 1.37444$ \times 10^{-25}$ & 0.00981745 \\
 \text{J0738-4042} & 41.1211 & 4.55418$ \times 10^{-24}$ & 0.309597 \\
 \text{J0742-2822} & 37.6139 & 8.28784$ \times 10^{-24}$ & 0.492893 \\
 \text{J0751+1807} & 74.4793 & 5.07004$ \times 10^{-25}$ & 0.0362146 \\
 \text{J0754+3231} & 38.44 & 2.40285$ \times 10^{-24}$ & 0.147635 \\
 \text{J0758-1528} & 49.8372 & 9.44338$ \times 10^{-25}$ & 0.0674527 \\
 \text{J0814+7429} & 42.932 & 1.27573$ \times 10^{-23}$ & 0.911233 \\
 \text{J0820-1350} & 41.3826 & 1.12452$ \times 10^{-23}$ & 0.77176 \\
 \text{J0823+0159} & 65.7034 & 7.55761$ \times 10^{-25}$ & 0.0539829 \\
 \text{J0826+2637} & 36.7006 & 3.31806$ \times 10^{-23}$ & 1.90188 \\
 \text{J0835-4510} & 48.0888 & 1.34572$ \times 10^{-23}$ & 0.961228 \\
 \text{J0837+0610} & 37.0345 & 1.56722$ \times 10^{-23}$ & 0.910601 \\
 \text{J0837-4135} & 39.0797 & 4.87755$ \times 10^{-24}$ & 0.307197 \\
 \text{J0846-3533} & 42.1711 & 2.1265$ \times 10^{-23}$ & 1.50133 \\
 \text{J0908-1739} & 38.8431 & 1.42168$ \times 10^{-23}$ & 0.887279 \\
 \text{J0922+0638} & 40.8917 & 2.12817$ \times 10^{-23}$ & 1.43466 \\
 \text{J0943+1631} & 36.5052 & 7.482$ \times 10^{-24}$ & 0.425441 \\
 \text{J0944-1354} & 50.0532 & 5.0256$ \times 10^{-24}$ & 0.358971 \\
 \text{J0946+0951} & 36.5382 & 1.31884$ \times 10^{-23}$ & 0.750931 \\
 \text{J0953+0755} & 64.7903 & 4.58178$ \times 10^{-24}$ & 0.32727 \\
 \text{J1012+5307} & 53.4332 & 5.4258$ \times 10^{-24}$ & 0.387557 \\
 \text{J1024-0719} & 38.7005 & 1.71988$ \times 10^{-23}$ & 1.06749 \\
 \text{J1041-1942} & 36.6277 & 4.21129$ \times 10^{-24}$ & 0.240668 \\
 \text{J1045-4509} & 37.5731 & 3.23577$ \times 10^{-24}$ & 0.192124 \\
 \text{J1115+5030} & 39.0036 & 1.62815$ \times 10^{-23}$ & 1.02244 \\
 \text{J1116-4122} & 45.2874 & 1.73472$ \times 10^{-24}$ & 0.123909 \\
 \text{J1136+1551} & 42.1747 & 8.4083$ \times 10^{-23} $& 5.9371 \\
 \text{J1239+2453} & 40.3084 & 2.84975$ \times 10^{-23}$ & 1.88014 \\
 \text{J1300+1240} & 35.5643 & 2.88182$ \times 10^{-23}$ & 1.57571 \\
 \text{J1321+8323} & 36.5293 & 1.66243$ \times 10^{-23}$ & 0.946228 \\
 \text{J1328-4357} & 41.7366 & 1.25324$ \times 10^{-23}$ & 0.871163 \\
 \text{J1430-6623} & 38.6252 & 1.02823$ \times 10^{-23}$ & 0.636333 \\
 \text{J1453-6413} & 37.0066 & 7.7727$ \times 10^{-24}$ & 0.451106 \\
 \text{J1455-3330} & 46.1529 & 5.94254$ \times 10^{-24}$ & 0.424467 \\
 \text{J1456-6843} & 45.4369 & 1.0231$ \times 10^{-23}$ & 0.730788 \\
 \text{J1509+5531} & 40.5816 & 2.3723$ \times 10^{-23}$ & 1.58107 \\
 \text{J1518+4904} & 67.6675 & 1.25824$ \times 10^{-24}$ & 0.0898742 \\
 \text{J2127+1155} & 42.3434 & 6.41853$ \times 10^{-24}$ & 0.455935 \\
 \text{J1543-0620} & 44.2132 & 4.34578$ \times 10^{-24}$ & 0.310413 \\
 \text{J1543+0929} & 43.8759 & 2.08026$ \times 10^{-24}$ & 0.14859 \\
 \text{J1559-4438} & 42.6042 & 3.56121$ \times 10^{-24}$ & 0.254372 \\
 \text{J1600-3053} & 46.2049 & 1.64159$ \times 10^{-24}$ & 0.117257 \\
 \text{J1603-7202} & 57.8922 & 1.28057$ \times 10^{-24}$ & 0.0914696 \\
 \text{J1604-4909} & 40.8763 & 7.19655$ \times 10^{-24}$ & 0.484865 \\
 \text{J1607-0032} & 73.4576 & 6.51078$ \times 10^{-25}$ & 0.0465056 \\
 \text{J1623-2631} & 38.4278 & 7.80249$ \times 10^{-24}$ & 0.479169 \\
 \text{J1640+2224} & 52.6765 & 2.48508$ \times 10^{-24}$ & 0.177506 \\
 \text{J1643-1224} & 36.5278 & 2.60936$ \times 10^{-24}$ & 0.148511 \\
 \text{J1645-0317} & 39.5595 & 7.80952$ \times 10^{-24}$ & 0.500944 \\
 \text{J1709-1640} & 74.3462 & 2.54331$ \times 10^{-25}$ & 0.0181665 \\
 \text{J1709+2313} & 45.5546 & 2.4967$ \times 10^{-24}$ & 0.178335 \\
 \text{J1713+0747} & 66.3388 & 9.09886$ \times 10^{-25}$ & 0.0649919 \\
 \text{J1720-0212} & 41.9078 & 5.9315$ \times 10^{-24}$ & 0.414854 \\
 \text{J1722-3207} & 41.6931 & 9.28191$ \times 10^{-24}$ & 0.644202 \\
 \text{J1735-0724} & 41.8263 & 6.49042$ \times 10^{-24}$ & 0.452622 \\
 \text{J1738+0333} & 52.7357 & 1.49566$ \times 10^{-24}$ & 0.106833 \\
 \text{J1740+1311} & 42.0383 & 6.73415$ \times 10^{-24}$ & 0.473193 \\
 \text{J1741-3927} & 40.304 & 5.13915$ \times 10^{-24}$ & 0.339001 \\
 \text{J1744-1134} & 59.8904 & 3.77654$ \times 10^{-24}$ & 0.269753 \\
 \text{J1745-3040} & 50.3561 & 1.63708$ \times 10^{-24}$ & 0.116934 \\
 \text{J1752-2806} & 60.1363 & 1.16259$ \times 10^{-24}$ & 0.0830422 \\
 \text{J1801-2451} & 48.0157 & 8.50308$ \times 10^{-25}$ & 0.0607363 \\
 \text{J1801-1417} & 36.5256 & 7.3598$ \times 10^{-24}$ & 0.418842 \\
 \text{J1803-2137} & 38.877 & 5.109$ \times 10^{-24}$ & 0.319274 \\
 \text{J1807-0847} & 45.9144 & 1.24016$ \times 10^{-24}$ & 0.0885829 \\
 \text{J1820-0427} & 37.5413 & 4.30552$ \times 10^{-24}$ & 0.255316 \\
 \text{J1823+0550} & 48.5757 & 1.25897$ \times 10^{-24}$ & 0.0899267 \\
 \text{J1824-1945} & 40.9223 & 2.43213$ \times 10^{-23}$ & 1.6414 \\
 \text{J1824-2452} & 42.3423 & 1.19895$ \times 10^{-24}$ & 0.085163 \\
 \text{J1825-0935} & 49.139 & 3.6633$ \times 10^{-24}$ & 0.261664 \\
 \text{J1829-1751} & 54.6043 & 2.63599$ \times 10^{-23}$ & 1.88285 \\
 \text{J1832-0827} & 40.1473 & 5.06412$ \times 10^{-24}$ & 0.332107 \\
 \text{J1833-0827} & 41.7718 & 7.73006$ \times 10^{-24}$ & 0.538018 \\
 \text{J1835-1106} & 41.7953 & 1.43077$ \times 10^{-23}$ & 0.996669 \\
 \text{J1836-0436} & 37.6347 & 3.43269$ \times 10^{-24}$ & 0.204318 \\
 \text{J1836-1008} & 41.3733 & 5.08732$ \times 10^{-24}$ & 0.349026 \\
 \text{J1840+5640} & 38.4105 & 1.00787$ \times 10^{-23}$ & 0.61854 \\
 \text{J1841-0425} & 38. & 3.26129$ \times 10^{-24}$ & 0.196949 \\
 \text{J1844+1454} & 40.4852 & 1.11881$ \times 10^{-23}$ & 0.743005 \\
 \text{J1850+1335} & 37.1645 & 4.66117$ \times 10^{-24}$ & 0.272255 \\
 \text{J1857+0943} & 69.9194 & 7.32693$ \times 10^{-25}$ & 0.0523352 \\
 \text{J1900-2600} & 40.5195 & 1.24926$ \times 10^{-23}$ & 0.830687 \\
 \text{J1902+0615} & 39.1075 & 2.16284$ \times 10^{-24}$ & 0.136365 \\
 \text{J1905-0056} & 38.4463 & 2.48948$ \times 10^{-24}$ & 0.152996 \\
 \text{J1906+0641} & 37.6858 & 1.73977$ \times 10^{-24}$ & 0.103764 \\
 \text{J1907+4002} & 39.9779 & 4.40831$ \times 10^{-24}$ & 0.28727 \\
 \text{J1909+1102} & 36.5478 & 2.89161$ \times 10^{-24}$ & 0.16471 \\
 \text{J1909-3744} & 36.5355 & 1.1301$ \times 10^{-23}$ & 0.643394 \\
 \text{J1910-5959A} & 44.7141 & 1.20734$ \times 10^{-24}$ & 0.0862383 \\
 \text{J1910-5959C} & 41.4165 & 1.64278$ \times 10^{-24}$ & 0.112883 \\
 \text{J1911-1114} & 36.8491 & 7.30991$ \times 10^{-24}$ & 0.421542 \\
 \text{J1913+1400} & 38.7526 & 3.7195$ \times 10^{-24}$ & 0.231327 \\
 \text{J1913-0440} & 39.8321 & 2.4511$ \times 10^{-24}$ & 0.158855 \\
 \text{J1915+1009} & 36.9264 & 2.15708$ \times 10^{-24}$ & 0.124784 \\
 \text{J1915+1606} & 45.6686 & 6.36111$ \times 10^{-25}$ & 0.0454365 \\
 \text{J1916+0951} & 40.7829 & 2.47226$ \times 10^{-24}$ & 0.165997 \\
 \text{J1917+1353} & 41.5323 & 1.59151$ \times 10^{-24}$ & 0.109819 \\
 \text{J1919+0021} & 65.5118 & 3.34207$ \times 10^{-25}$ & 0.0238719 \\
 \text{J1921+2212} & 41.7992 & 9.49871$ \times 10^{-24}$ & 0.661766 \\
 \text{J1926+1648} & 57.7089 & 3.11667$ \times 10^{-24}$ & 0.222619 \\
 \text{J1932+1059} & 36.9246 & 3.18859$ \times 10^{-23}$ & 1.84443 \\
 \text{J1935+1616} & 40.6174 & 3.16151$ \times 10^{-24}$ & 0.210985 \\
 \text{J1937+2544} & 37.0187 & 5.19088$ \times 10^{-24}$ & 0.301413 \\
 \text{J1939+2134} & 76.5964 & 5.27885$ \times 10^{-26}$ & 0.0037706 \\
 \text{J1941-2602} & 38.8781 & 4.24401$ \times 10^{-24}$ & 0.26523 \\
 \text{J1944+0907} & 39.8356 & 6.16328$ \times 10^{-24}$ & 0.399491 \\
 \text{J1946-2913} & 42.4317 & 8.50239$ \times 10^{-24}$ & 0.60585 \\
 \text{J1946+1805} & 64.8701 & 1.3907$ \times 10^{-24}$ & 0.099336 \\
 \text{J1948+3540} & 40.2965 & 3.1063$ \times 10^{-24}$ & 0.204849 \\
 \text{J1952+3252} & 38.3669 & 6.99324$ \times 10^{-24}$ & 0.428452 \\
 \text{J1954+2923} & 45.8579 & 1.06684$ \times 10^{-23}$ & 0.762031 \\
 \text{J1955+2908} & 43.7264 & 9.59627$ \times 10^{-25}$ & 0.0685448 \\
 \text{J1955+5059} & 40.8668 & 1.40886$ \times 10^{-23}$ & 0.948881 \\
 \text{J1959+2048} & 36.7592 & 9.00599$ \times 10^{-24}$ & 0.517451 \\
 \text{J2002+4050} & 41.0534 & 3.17817$ \times 10^{-24}$ & 0.215521 \\
 \text{J2004+3137} & 40.3195 & 2.74782$ \times 10^{-24}$ & 0.181364 \\
 \text{J2013+3845} & 35.0351 & 1.2161$ \times 10^{-23}$ & 0.650212 \\
 \text{J2018+2839} & 67.6517 & 9.09648$ \times 10^{-25}$ & 0.0649749 \\
 \text{J2019+2425} & 43.7666 & 5.66758$ \times 10^{-24}$ & 0.404827 \\
 \text{J2022+2854} & 38.4464 & 6.63234$ \times 10^{-24}$ & 0.407605 \\
 \text{J2022+5154} & 41.1168 & 3.41183$ \times 10^{-24}$ & 0.231902 \\
 \text{J2023+5037} & 36.6147 & 6.76838$ \times 10^{-24}$ & 0.386596 \\
 \text{J2046-0421} & 36.5212 & 3.44972$ \times 10^{-24}$ & 0.196239 \\
 \text{J2046+1540} & 37.6523 & 4.05443$ \times 10^{-24}$ & 0.241493 \\
 \text{J2048-1616} & 38.8761 & 3.17922$ \times 10^{-23}$ & 1.9867 \\
 \text{J2051-0827} & 67.0076 & 7.40902$ \times 10^{-25}$ & 0.0529216 \\
 \text{J2055+2209} & 57.4892 & 1.00494$ \times 10^{-24}$ & 0.0717813 \\
 \text{J2055+3630} & 41.9797 & 1.10321$ \times 10^{-24}$ & 0.0773579 \\
 \text{J2108+4441} & 44.3737 & 9.36083$ \times 10^{-25}$ & 0.0668631 \\
 \text{J2113+2754} & 39.1337 & 1.56216$ \times 10^{-23}$ & 0.985915 \\
 \text{J2113+4644} & 38.4549 & 3.59792$ \times 10^{-24}$ & 0.221191 \\
 \text{J2116+1414} & 37.9948 & 3.80432$ \times 10^{-24} $& 0.229695 \\
 \text{J2124-3358} & 53.7137 & 1.10886$ \times 10^{-23}$ & 0.792045 \\
 \text{J2129+1210A} & 36.6794 & 1.3541$ \times 10^{-24}$ & 0.0775484 \\
 \text{J2129+1210B} & 42.2722 & 6.54473$ \times 10^{-25}$ & 0.0463727 \\
 \text{J2129+1210C} & 37.31 & 1.08497$ \times 10^{-24}$ & 0.0637445 \\
 \text{J2129-5721} & 43.8083 & 2.01653$ \times 10^{-24}$ & 0.144038 \\
 \text{J2145-0750} & 66.379 & 2.02952$ \times 10^{-24}$ & 0.144966 \\
 \text{J2149+6329} & 40.4712 & 4.24225$ \times 10^{-24}$ & 0.281582 \\
 \text{J2150+5247} & 36.9824 & 2.51365$ \times 10^{-24}$ & 0.145743 \\
 \text{J2157+4017} & 40.3436 & 4.42204$ \times 10^{-24}$ & 0.292128 \\
 \text{J2219+4754} & 39.1588 & 8.63068$ \times 10^{-24}$ & 0.545228 \\
 \text{J2225+6535} & 41.5695 & 4.25672$ \times 10^{-23}$ & 2.94121 \\
 \text{J2229+6205} & 37.6982 & 2.82253$ \times 10^{-24}$ & 0.168426 \\
 \text{J2229+2643} & 41.5673 & 4.51013$ \times 10^{-24}$ & 0.311606 \\
 \text{J2235+1506} & 43.6668 & 4.51705$ \times 10^{-24}$ & 0.322647 \\
 \text{J2257+5909} & 41.3007 & 4.20562$ \times 10^{-24}$ & 0.287776 \\
 \text{J2305+3100} & 39.1606 & 5.3719$ \times 10^{-24}$ & 0.339384 \\
 \text{J2308+5547} & 37.1242 & 4.60111$ \times 10^{-24}$ & 0.26831 \\
 \text{J2313+4253} & 43.4977 & 5.53556$ \times 10^{-24}$ & 0.395397 \\
 \text{J2317+1439} & 51.393 & 1.6924$ \times 10^{-24}$ & 0.120886 \\
 \text{J2321+6024} & 37.8303 & 5.16906$ \times 10^{-24}$ & 0.31007 \\
 \text{J2322+2057} & 44.9277 & 6.10354$ \times 10^{-24}$ & 0.435967 \\
 \text{J2326+6113} & 39.7623 & 4.82672$ \times 10^{-24}$ & 0.311996 \\
 \text{J2330-2005} & 37.0468 & 2.29858$ \times 10^{-23}$ & 1.33621 \\
 \text{J2337+6151} & 40.4429 & 2.95763$ \times 10^{-24}$ & 0.196108 \\
 \text{J2354+6155} & 38.8978 & 6.1822$ \times 10^{-24}$ & 0.386652\\
\hline \hline
\end{longtable}



\section{Computing detectability of GW from RAPs}

{ One can discuss the detectability of the GW (burst) signal of the rise fling 
of galactic RAPs by using the {\sl matching filter} technique. This procedure 
is based on the definition of the {\sl signal-to-noise ratio} (SNR), which is 
a quantity used as a criterium of detectability of a GW signal. The SNR 
depends on the features in the GW waveform of the process, the 
orientation of the source with respect to the GW interferometer (observer) and 
also on the source direction. It also depends on the total energy per unit 
frequency, $dE/df(f)$ (the GW energy spectrum at the source), that is carried 
away from the source by the GW, and on the distance to the source $D_\star$. Here 
$f$ is the average GW frequency for a given signal. Because of this angular 
dependence, the SNR is usually defined as 

\begin{equation}
SNR = \langle \rho^{2} \rangle = \frac{2 }{5 \pi^{2} D_\star } \int^{\infty}_{0} 
\frac{1}{f^{2} S_{h} \left(f\right)} \frac{dE}{df} ~ f ~df \; .
\label{6a}     
\end{equation}

Symbol $\langle \rangle$ refers to an average of SNR square for all the directions 
(angles) to the GW source. In other words, it is obtained from the average $rms$ 
of the amplitudes of the GW signals for different orientation of the source and the 
interferometer. In our computations above in Section III we applied this definition 
to estimate the S/N ratio given in Table II. As the attentive reader may notice, in 
what follows we shall use an equivalent definition of the SNR to quantify the 
detectability of RAPs GW signals in the context of the formalism of Ref.\cite{ori2001}
}

To complete this brief discussion on the SNR, we collect below the power-laws
that according to most of the experimental studies with GW interferometers are 
what better describe the interferometer noise in each band of the spectrum. To 
this purpose, it is defined the dimensionless quantity: $h_{rms}(f)\equiv 
\sqrt{f S_{h}(f)}$, wherein $S_{h}(f)$ is the interferometer noise spectral 
density. Thus, the strain sensitivity for each frequency band is given by

\begin{equation}
h_{rms}\left(f\right)=
\left\{\begin{array}{ll}
\infty, & f < f_{s},\\
h_{m}\left(\alpha f/f_{m}\right)^{-3/2}, & f_{s}\leqslant f < f_{m}/\alpha,\\
h_{m}, & f_{m}/\alpha \leqslant f < \alpha f_{m}, \\
h_{m}\left[f/\left(\alpha f_{m}\right)\right]^{3/2}, & \alpha f_{m} < f \; \; .
\end{array}\right.
\label{7a}
\end{equation}

Understanding these laws is straightforward if one recalls that the interferometer 
noise curves depend basically on four parameters:

\begin{itemize}
\item The threshold frequency $f_{s}$: below of it the noise grows almost 
asymptotically. For detectors on Earth the noise source is the seismic gradients

\item The frequency $f_{m}$, centered in the almost flat part of the noise spectrum 

\item A critical amplitude $h_{m}$, which is the minimum value of $h_{rms}(f)$

\item And the dimensionless parameter $\alpha$, which determines the width of the flat 
band of the noise curve 
\end{itemize}
 
For instance, for Advanced LIGO the following set of parameters is estimated (see E. 
E. Flanagan and S. A. Hughes, [Phys. Rev. D 57, 8 (1998)]) 

\begin{equation}
\text{LIGO Advanced} \left\{ \begin{array}{l}
f_{s}=10~{\rm Hz}\\
f_{m}=68~ {\rm Hz}\\
\alpha=1.6 \\
h_{m}= 1.4 \times 10^{-23}
\end{array}
\right. \; .
\label{8a}
\end{equation}


\begin{figure*}[tbh]
\centering
\subfigure[\label{k=t} LIGO I, LIGO Advanced, TAMA, VIRGO and GEO600 
strain sensitivities and the $h_{c}$ and $f_{c}$ characteristics of 
the GW signal produced by each of the 212 pulsars calibrated with 
respect to LIGO Advanced, for parameters $a=10^4$ km s$^{-2}$ and 
$V_{\rm kick}\sim$ 10$^4$ km s$^{-1}$. ]
{\includegraphics[height=3.0in,width=3.0in]{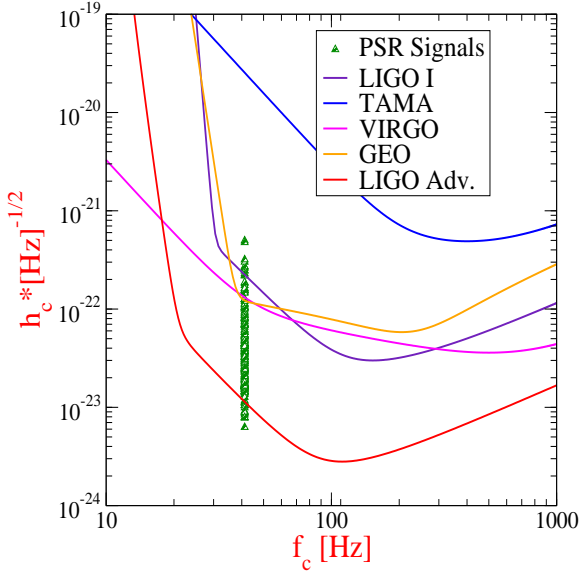} } 
\hskip 1.5truecm
\subfigure[\label{k=u} Idem as Fig.(\ref{k=t}), but now for parameters 
$a=10^6$ km s$^{-2}$ and $V_{\rm kick}\sim$ 10$^4$ km s$^{-1}$, calibrated with 
respect to Advanced LIGO. It is possible to see that these signals have 
high enough chance to be detected. ]{
\includegraphics[height=3.0in,width=3.0in]{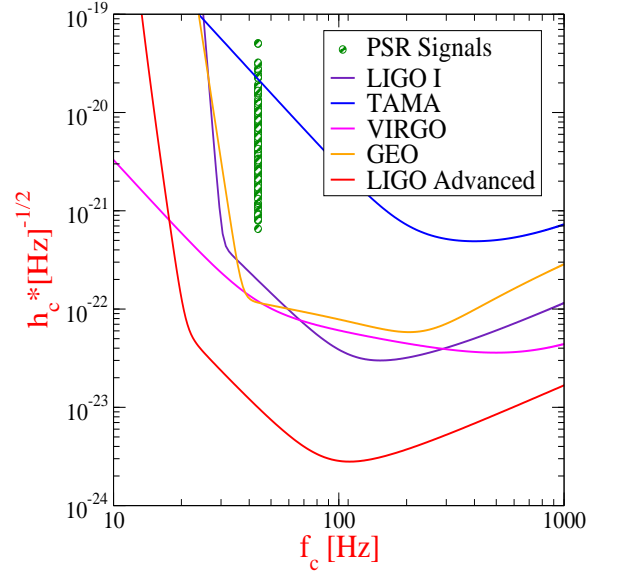} }
\caption{Color Online. Pulsar GW signals for different parameters $a$ and 
$V_{\rm kick}$.  (a), (b)} 
\label{figure3}
\end{figure*}

\begin{figure}[tbh]
\centering
\subfigure[\label{k=v} Idem as Fig.(\ref{k=t}), but now for parameters  
$a=10^6$ km s$^{-2}$ and  $V_{\rm kick} \sim$ 10$^4$ km s$^{-1}$, and calibrated 
against LIGO I. Clearly also these signals might be detected ]{
\includegraphics[height=3.0in,width=3.0in]{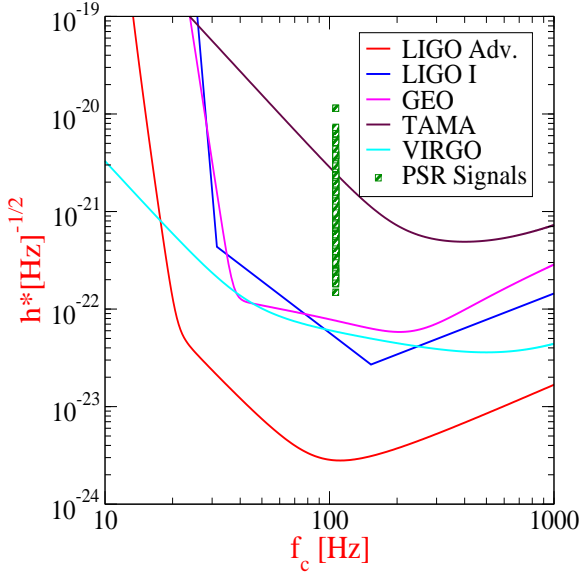} }
\caption{Color Online. Pulsar GW signals for different parameters $a$ and $V$. (c) } 
\label{CALIB-LIGO-II}
\end{figure}

Therefore, to know whether the RAPs GW signals can be effectively detected
by the currently operative interferometric detectors LIGO, VIRGO, 
GEO-600, etc., one needs to compute the signal-to-noise ratio with 
respect to the strain sensitivity of these instruments. (We refer 
to this procedure as {\sl calibration}). In order to do that we need 
first to compute the frequency at which the largest part 
of the GW emission is expected to take place. This is the so-called 
characteristic frequency $f_c$, which is dependent on the detector 
design properties. 

Should the detector have a power spectral density $S_h(f)$, have the 
characteristic frequency to  be obtained from the frequency first 
moment of the time-dependent GW amplitude $h(t)$, i. e.,

\begin{equation}
f_c \equiv \left(\int_0^\infty \frac{\langle|\tilde{h}(f)|^2 
\rangle} {S_h(f)}~ f df \right) \left[\int_0^\infty 
\frac{\langle|\tilde{h}(f)|^2 \rangle} {S_h(f)}~df \right]^{-1} ,
\end{equation}

where $\langle \tilde{h}(f)\rangle$ is the Fourier transform of $h(t)$
\footnote{The Fourier transform reads: $ \tilde{h} = 
\int^{\infty}_{-\infty} e^{2\pi ift} h\left(t\right)dt $ }, and the 
symbol $\langle \rangle$ stands for an 
average over randomly distributed angles of $h(f)$, and it can be 
approximated as $\langle |\tilde{h}(f)|\rangle^2 =  |\tilde{h}(f)|^2$. 

The characteristic frequency $f_c$ is strictly related to the characteristic 
amplitude $h_c$, which is the physical quantity associated to the GW pulse 
$h(t)$ to be detected by a given instrument. It can be computed as 

\begin{equation}
h_c(f_c) \equiv \left(3 \int_0^\infty \frac{S_h(f_c)}{S_h(f)} 
\langle|\tilde{h}(f)|^2 \rangle ~f df \right)^{1/2} .
\end{equation}

In this way we take $\tilde{h}$ in the following form (see Ref.\cite{sago04, 
ori2001})

\begin{equation}
\vert \tilde{h} \vert^{2} = \frac{\left(\Delta h_{m}\right)}{8\pi^{4}f^{4} 
t^{2}_{m}} \left(1 - \cos(2\pi t_{m} f) \right) \, .
\label{10a} 
\end{equation}

{ Here $t_{m}$ defines the rise time of the memory of the signal. This equation 
is obtained from the Eq.(\ref{hmax}) given above. Thus, Eq.(\ref{10a}) is the 
Fourier transform of the GW waveform, where $\Delta h$ is given by (after average
over angles in Eq.(\ref{hmax})) }

\begin{equation}
\Delta h_{m} = \frac{4G \beta^{2} \gamma M_{\star}}{c^{2}D_{\star}}
\label{11a}
\end{equation}

with $G$ the universal constant of gravitation, $\gamma$ Lorentz factor, 
$M_{\star}$ star mass, $D_{\star}$  distance to Earth, and $\beta^2 = 
|\vec{V}|^{2}/c^{2}$.



Therefore, the signal-to-noise ratio $S/N$ is now computed using the relation

\begin{equation}
\frac{S}{N} = \frac{h_{c}(f_c)}{h_{rms}\left(f_{c}\right)}
\label{13a}
\end{equation}

{ where $h_{rms}\left(f_{c}\right) = \left[f_{c}S_{h}\left(f_{c}\right) \right]^{1/2}$  
is the average amplitude of the interferometer spectral noise at the 
characteristic frequency.}


Guided by Table-\ref{sn-ejecta}, this allows us to compute the characteristic 
amplitude of the GW burst released during the short rise time of acceleration 
of each RAP for which those parameters are estimated. Since we are using only 
the transversal velocity of the pulsars, then we set $\theta \simeq 90^0$ in 
our calculations below. Our main results are presented in Figs.(\ref{k=t}, 
\ref{k=u}, \ref{k=v}). Those Figures compare the GW signals from individual 
RAPs with the expected sensitivities of LIGO I, its projected advanced 
configuration LIGO II, VIRGO, GEO-600, and TAMA-300 observatories.  

{
At this stage, a note of caution to the attentive reader is needed. It regards 
with the appearance of the noise curves of TAMA, GEO and VIRGO GW observatories
in figures from 4 through 9, in addition to the noise curves for LIGO I and LIGO 
Advanced, despite that the calibration of the expected signals in those figures 
(symbol: triangles, 
circles, etc.) is performed exclusively with respect to LIGO I and LIGO Advanced, 
independently. First, the reader must keep in mind that the position of those 
points in the diagram h$_c$ vs. f$_c$ will change when the calibration of the 
expect GW strain is performed 
with respect to the noise curve of a different detector, say TAMA, VIRGO, etc.. 
For instance, compare Fig.4(a) and Fig.5(a) which are calibrated for LIGO Advanced 
and LIGO I, respectively. Second, thinking on the benefit of the Scientific Community 
operating observatories like TAMA, GEO and VIRGO, those noise curves are included having
in mind the expectation that such detectors might gain more sensitivity in both frequency 
band and amplitude so as to be able to catch the GW signals produced by Milky Way RAPs.

Finally, but not to the end, a direct interpretation of all the h$_c$ vs. f$_c$ diagrams 
in figures from 4 through 9 is that, exception done for the TAMA-300 GW detector, the 
signals (points) plotted in those figures appear to be not very much sensitive to the 
noise curves of the instruments LIGO I, LIGO Advanced, VIRGO and GEO-600. This feature 
justifies to include those noise curves in such figures.
}

\begin{figure*}[tbh]
\centering
\subfigure[\label{L2eff=0.4} Efficiency case $\epsilon = 0.40$!! for parameters  
$a = 10^6$km s$^{-2}$ and $V_{\rm kick} \sim$ 10$^4$ km s$^{-1}$, calibrated 
against LIGO Advanced. ]{
\includegraphics[height=3.0in,width=3.0in]{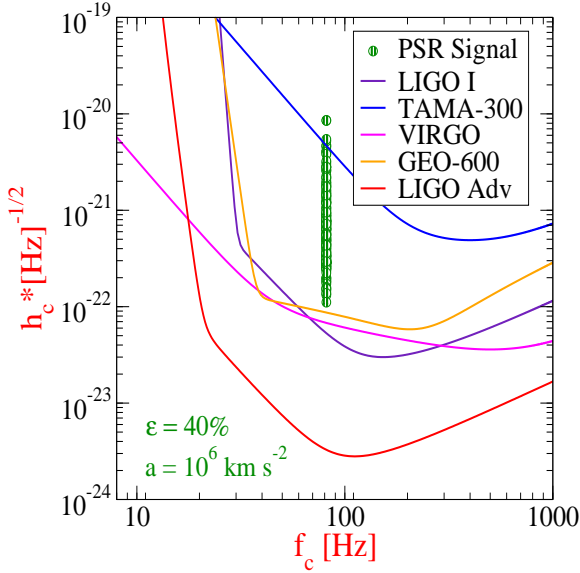} }
\hskip 0.5 truecm
\subfigure[\label{L2eff=0.1} Efficiency case $\epsilon = 0.10$!! for parameters  
$a = 10^6$ km s$^{-2}$ and $V_{\rm kick} \sim$ 10$^4$ km s$^{-1}$, calibrated 
against LIGO Advanced. ]{
\includegraphics[height=3.0in,width=3.0in]{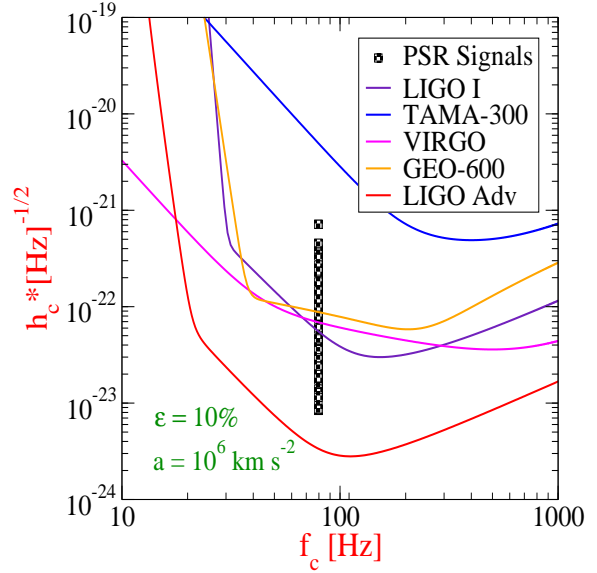} }
{\vskip 1.0truecm
\subfigure[\label{L2eff=0.05} Efficiency case $\epsilon = 0.05$!! for parameters  
$a = 10^6$ km s$^{-2}$ and $V_{\rm kick} \sim$ 10$^4$ km s$^{-1}$, calibrated 
against LIGO Advanced. ]{
\includegraphics[height=3.0in,width=3.0in]{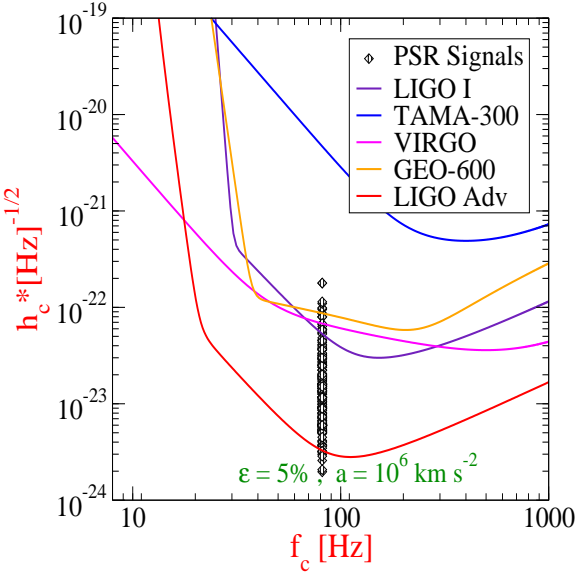} }
\hskip 0.5truecm
\subfigure[\label{L2eff=0.01} Efficiency case $\epsilon = 0.01$!! for parameters  
$a = 10^6$km s$^{-2}$ and $V_{\rm kick} \sim$ 10$^4$ km s$^{-1}$, calibrated 
against LIGO Advanced. ]{
\includegraphics[height=3.0in,width=3.0in]{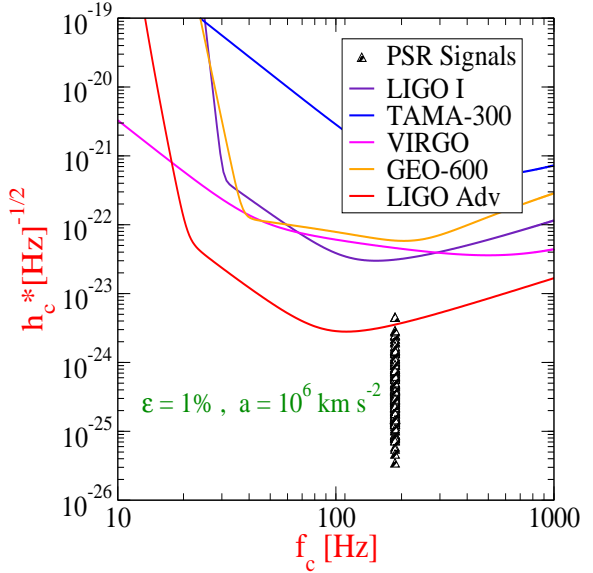} }
}
\caption{Color Online. Pulsar GW signals for parameters: $a = 10^4$ km s$^{-2}$ and 
$V_{\rm kick} \sim$ 10$^4$ km s$^{-1}$} 
\label{L2efficiency-a6}
\end{figure*}

\section{More realistic analysis of GW emission from RAPs}

We can start with by recalling both the inertia (which is large for a neutron 
star progenitor) and linear momentum conservation laws, which should dominate 
the physics of the SN explosion that gives the kick to the neutron star. Thus, 
one can write

\begin{equation}
M_{SN} V_{SN} = M_{NS} V_{NS} - M_{RSN} V_{RSN} \; ,
\label{momento}
\end{equation}

where M$_{SN}$, M$_{NS}$, and M$_{RSN}$ are the SN, neutron star, and ejected 
material mass, respectively. Be aware of the role of the ``$-$'' signal in front 
of the term for the ejected material. It is responsible for a correct description 
of the physics at the explosion. Without it, one would obtain immaginary values 
for the searched parameter $V_{NS}$. Now, by assuming parameters of a typical 
progenitor star M$_{SN}=10$ M$_{\odot}$, $V_{SN}=15$ km s$^{-1}$, M$_{NS}=1.4$ 
M$_{\odot}$ for a canonical neutron star, and for the material in the ejected 
envelope M$_{RSN} = 8.6 M_{\odot}$, and expansion velocity $V_{RSN} = 3000$ km 
s$^{-1}$ (see Table-\ref{sn-ejecta}), one obtains that the kick at birth velocity 
reads: $V_{0} = V_{\rm kick} \sim 10^{4}$ km s$^{-1}$ (compare to velocities taken 
from ATNF in Fig.\ref{PSR-VELOC-DIST}). The discrepancy between them will be briefly 
discussed below.

\subsection{Case a: Centered kick $V = V_{\rm kick}$ and $\epsilon = 1$ }

We first consider that the whole energy obtained from the conservation law 
discussed above goes into the pulsar kick, imparting to it no rotation at all.
This is a very idealistic situation. Nonetheless, it gives us an idea on the 
potential detectability of those GW signals from the flinging of a pulsar out 
of the SN cocoon. In other words, in this first approach we assume that the 
efficiency, $\epsilon$, that relates the effective RAP velocity $V$ (which 
enters the calculation of $h_c$) and the kick speed $V_{\rm kick}$ is  
$\epsilon = 1$! (In the other cases analyzed below these velocities will 
be related through the expression: $V = \epsilon V_{\rm kick}$.

\begin{table}
\caption{ATNF sample of pulsars having 2-Dim speed greater than $10^3$ 
km s$^{-1}$. The current high velocities suggest that a much higher speed 
was imparted at the kick at birth.}
\begin{center}
\begin{tabular}{lll}
\hline \hline
Jname & DIST &  VTrans \\
{ } & [kpc] & [km s$^{-1}$]\\
\hline\hline
\text{J0525+1115} &  7.68  & 1101.985 \\
\text{J1509+5531} &  2.41  & 1104.013 \\
\text{J1824-1945} &  5.20  & 2482.989 \\
\text{J1829-1751} &  5.49  & 3945.954 \\
\text{J2013+3845} &  13.07 & 2521.130 \\
\text{J2149+6329} &  13.65 & 1113.385 \\
\text{J2225+6535} &  2.00  & 1729.770 \\
\text{B2011+38}   &  ????  & 1624.0  \\
\text{B2224+65}   &  ????  & 1608.0  \\
\text{B1830-1945} &  5.20  & 740.0 \\
\text{$^\star$RX J0822-4300\footnotetext{$^\star$Ref. P. F. Winkler \& R. Petre, 
{\sl Direct measurement of neutron star recoil in the Oxygen-rich supernova remnant 
Puppis A}, Ap.J. 670, 635 (2007) } } & 2.0  & 1600.0 \\
\hline \hline
\label{pulsares}
\end{tabular}
\end{center}
\end{table}


\begin{table*}
\caption{	
High expansion speeds of CaII and SiII lines of the ejected material in some supernovae. 
Data taken from A. Balastegui, P. Ruiz-Lapuente, J. M\'endez, G. Altavilla, M. Irwin, K. 
Schahmaneche, C. Balland, R. Pain and N. Walton. arXiv: astro-ph/0502398v1 21 Feb 2005 }
\begin{center}
\begin{tabular}{lccc}
\hline \hline
Event & CaII Expansion Speed  & SiII Expansion Speed  & Days after maximum \\
{ } & [km s$^{-1}]$ & [km s$^{-1}$] & [days]\\
\hline\hline
\text{SN 2002li} &  17200$\pm$ 500  &                & -7\\
\text{SN 2002lj} &  12300$\pm$ 300  & 8900$\pm$ 300  &  7\\
\text{SN 2002lp} &  13800$\pm$ 900  & 10400$\pm$ 400 &  3\\
\text{SN 2002lq} &  18900$\pm$ 1400 &                & -7\\
\text{SN 2002lr} &  12800$\pm$ 300  &  8600$\pm$ 400 & 10\\
\text{SN 2002lk} &                  & 14500$\pm$ 500 & -5\\
\hline \hline
\label{sn-ejecta}
\end{tabular}
\end{center}
\end{table*}

\subsubsection{Fling with $a = 10^4$ {\rm km s}$^{-2}$, $V_{\rm kick}$ = 
$10^4$ {\rm km s}$^{-1}$}

Based on the dynamical analysis presented in the Section-\ref{psr-galactic-dynamics}, 
we suppose here that the short rise time kick imparts to the pulsar an acceleration 
$a= 10^4$ km s$^{-2}$, and a birth velocity $V_{\rm kick} = 10^4$ km s$^{-1}$. 
The GW signal obtained for these parameters is presented in Figs.(\ref{k=t}, \ref{k=u}).

\subsubsection{Fling with $a = 10^6$ {\rm km s}$^{-2}$, $V_{\rm kick}$ = 
$10^4$ {\rm km s}$^{-1}$}

Now one can take a step further and consider that the birth acceleration is actually 
more larger than the one currently inferred from the RAPs galactic parameters. Several 
physical mechanisms can act together to impart an acceleration as large as $a
= 10^6$ km s$^{-2}$, see for instance Refs.\cite{dong-lai2001,janka-muller-review,fryer2006}, 
and references therein. Such a large thrust can be estimated as $a\simeq V_
{kick}/\Delta T_{mech}$, where ($10^{-1} \lesssim \Delta T_{mech} \lesssim 10^{-3}$) s, 
depending on the specific mechanism driving the fling. The GW signal obtained for these 
parameters is presented in Fig.(\ref{k=v}).


\subsection{Case b: Off-centered kick}

\begin{figure*}[tbh]
\centering
\subfigure[\label{L2eff1=0.4} Efficiency case $\epsilon = 0.40$!! for parameters  
$a = 10^4$ km s$^{-2}$ and $V_{\rm kick} \sim$ 10$^4$ km s$^{-1}$, calibrated 
against LIGO Advanced. ]{
\includegraphics[height=3.0in,width=3.0in]{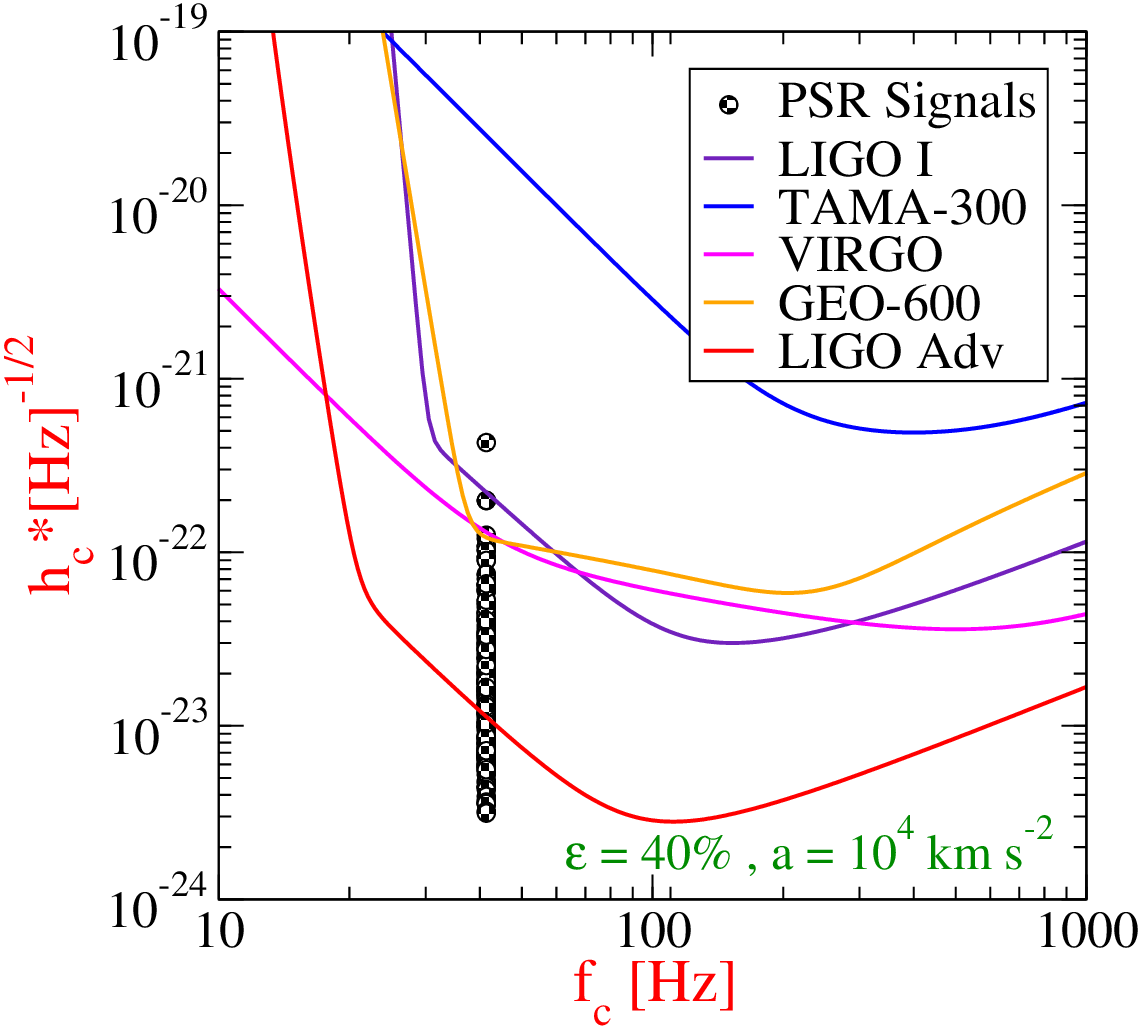} }
\hskip 0.5 truecm
\subfigure[\label{L2eff1=0.1} Efficiency case $\epsilon = 0.10$!! for parameters  
$a = 10^4$ km s$^{-2}$ and $V_{\rm kick} \sim$ 10$^4$ km s$^{-1}$, calibrated 
against LIGO Advanced. ]{
\includegraphics[height=3.0in,width=3.0in]{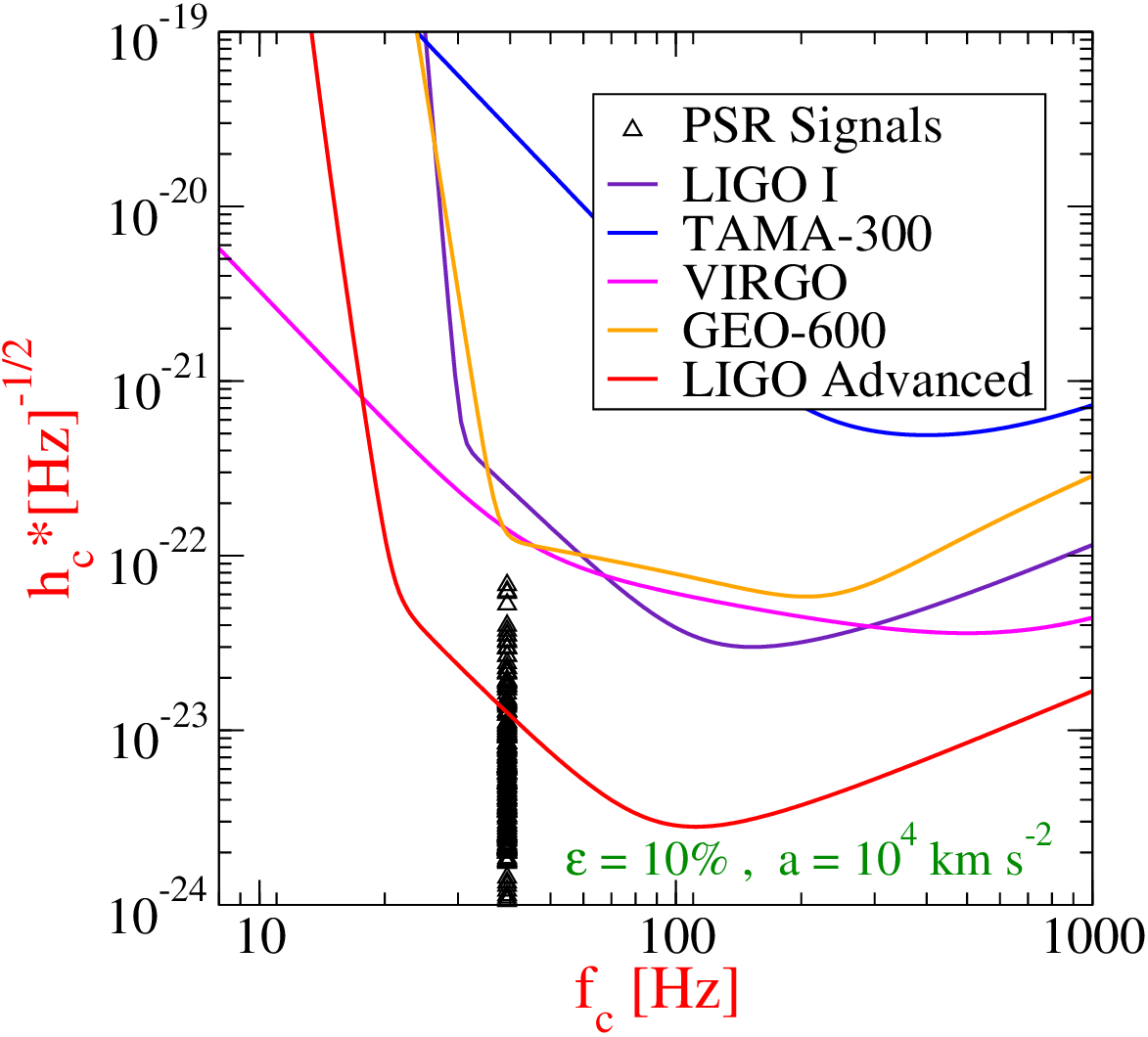} }
{\vskip 1.0truecm
\subfigure[\label{L2eff1=0.05} Efficiency case $\epsilon = 0.05$!! for parameters  
$a = 10^4$ km s$^{-2}$ and $V_{\rm kick} \sim$ 10$^4$ km s$^{-1}$, calibrated 
against LIGO Advanced. ]{
\includegraphics[height=3.0in,width=3.0in]{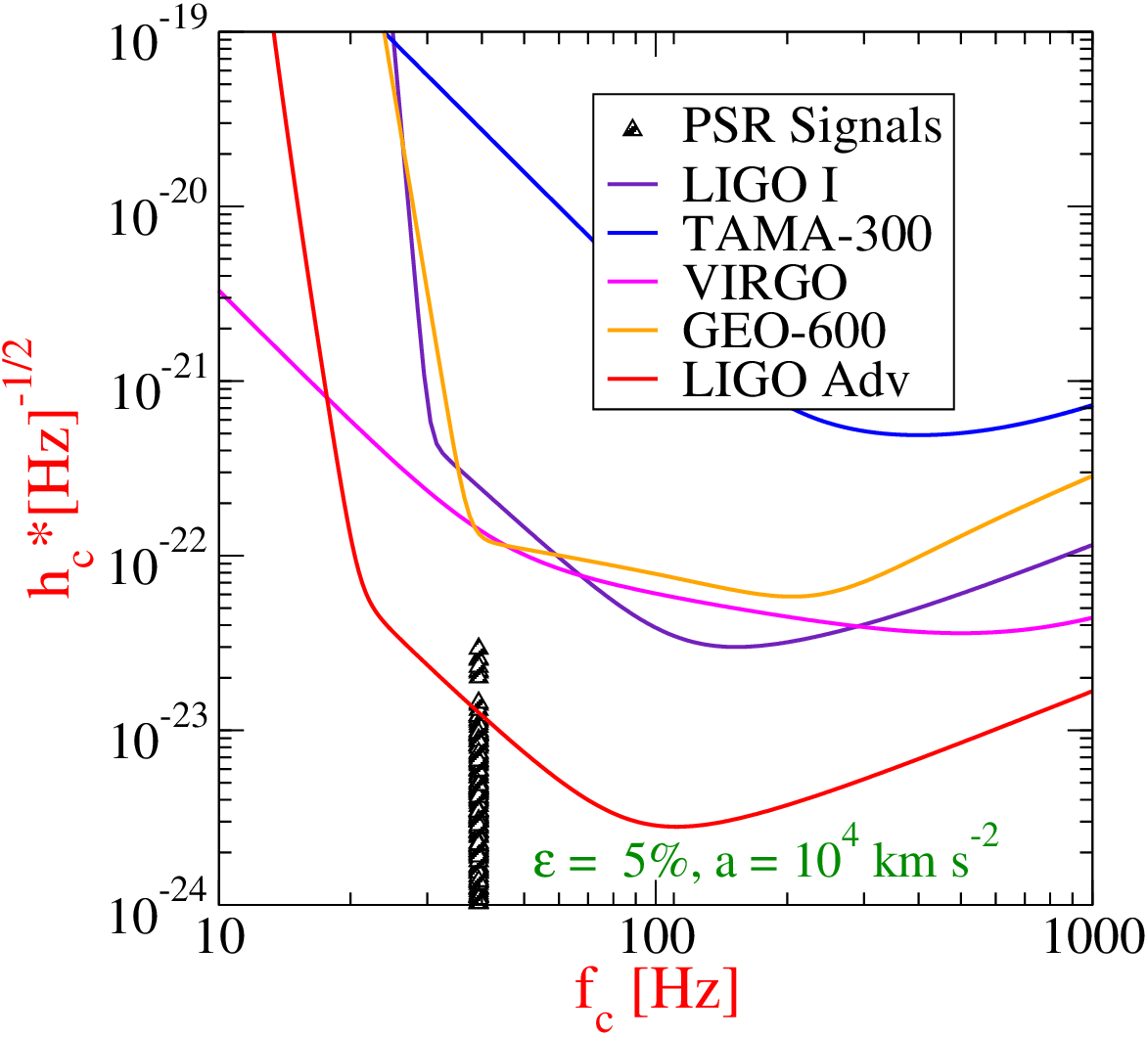} }
\hskip 0.5truecm
\subfigure[\label{L2eff1=0.01} Efficiency case $\epsilon = 0.01$!! for parameters  
$a = 10^4$ km s$^{-2}$ and $V_{\rm kick} \sim$ 10$^4$ km s$^{-1}$, calibrated
against LIGO Advanced. ]{
\includegraphics[height=3.0in,width=3.0in]{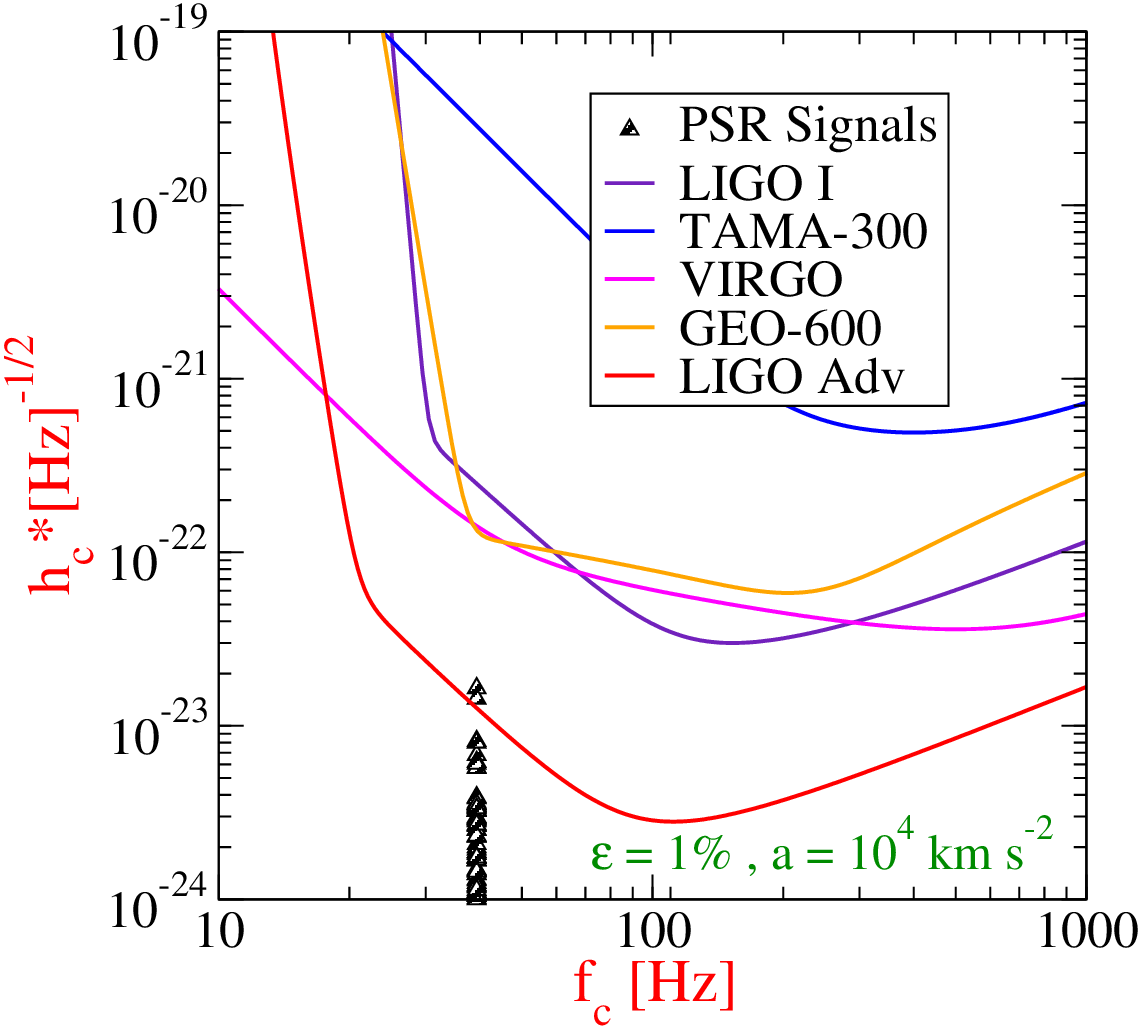} }
}
\caption{Color Online. Pulsar GW signals for parameters: $a = 10^4$ km s$^{-2}$ and 
$V_{\rm kick} \sim$ 10$^4$ km s$^{-1}$ } 
\label{L2efficiency}
\end{figure*}

Finally, we analyze the most astrophysically realistic kicks at birth: a situation where
the pulsar receives both a translational and a rotationl kick. In this case, the energy 
transferred to the translational kick can be parameterized by an efficiency $\epsilon$,
which ranges from 0 to 1. Of course, this efficiency represents our difficulty in explaining 
how a specific kick mechanism becomes the kick driver. {The remaining energy is supposed to 
be expended in making the pulsar to spin. This parametrization can be translated into a 
power-law relationship between the GW amplitude, $h$, and the efficiency, $\epsilon$, of 
energy conversion into pulsar kick, which has the form:

\begin{equation}
h \propto \epsilon^2 \, .
\label{h-vs-e}
\end{equation}

This relation stems from the quadratic dependence of the GW amplitude with the $\beta$ 
parameter 
appearing in Eq.(\ref{hmax}) above. } In this way, one can recover the whole spectrum of RAPs 
velocities as observed today and listed in ATNF catalog. To exemplify, we compute signals for 
values of $\epsilon = 0.01, 0.05, 0.10$, and $0.40$, which can reproduce the set of velocities 
$V_{\rm kick} = 100$ km s$^{-1}$, $V_{\rm kick} = 500$ km s$^{-1}$, $V_{\rm kick} = 10^3$ km 
s$^{-1}$, and $V_{\rm kick} = 4 \times 10^3$ km s$^{-1}$, as shown in 
Fig.-(\ref{PSR-VELOC-DIST}).
For this case, the results are presented in Figs.(\ref{L2efficiency-a6}, \ref{L2efficiency}, 
\ref{L1EFFICIENCY-a4}, \ref{EFF=CALIB-LIGO-I}). To the end of the paper a brief discussion on 
the pulsar periods provided by the ATNF catalog is given.


\subsubsection{Efficiency $\epsilon$ = 0.05}
The following figures illustrate the results obtained for this efficiency:
Fig.(\ref{L2eff1=0.05}) with $a = 10^6$ km s$^{-2}$ and Fig.(\ref{L2eff=0.05}) with 
$a = 10^4$ km s$^{-2}$ for advanced. Fig.(\ref{L1-eff=0.05}) with $a = 10^6$ km 
s$^{-2}$ and Fig.(\ref{L1eff1=0.4c}) with $a = 10^4$ km s$^{-2}$ para LIGO I.

\subsubsection{Efficiency $\epsilon$ = 0.10}
The following figures illustrate the results obtained for this efficiency:
Fig.(\ref{L2eff1=0.1}) with $a = 10^6$ km s$^{-2}$ and Fig.(\ref{L2eff=0.1}) with 
$a = 10^4$ km s$^{-2}$ for Advanced LIGO. Fig.(\ref{L1eff1=0.4b}) with $a = 10^6$ km 
s$^{-2}$ and Fig.(\ref{L1-eff=0.1}) with $a = 10^4$ km s$^{-2}$ for LIGO I.

\subsubsection{Efficiency $\epsilon$ = 0.40}
\begin{figure*}[tbh]
\centering
\subfigure[\label{L1-eff=0.4} Efficiency case $\epsilon = 0.40$!! for parameters  
$a = 10^4$ km s$^{-2}$ and $V_{\rm kick} \sim$ 10$^4$ km s$^{-1}$, calibrated 
against LIGO I. ]{
\includegraphics[height=3.0in,width=3.0in]{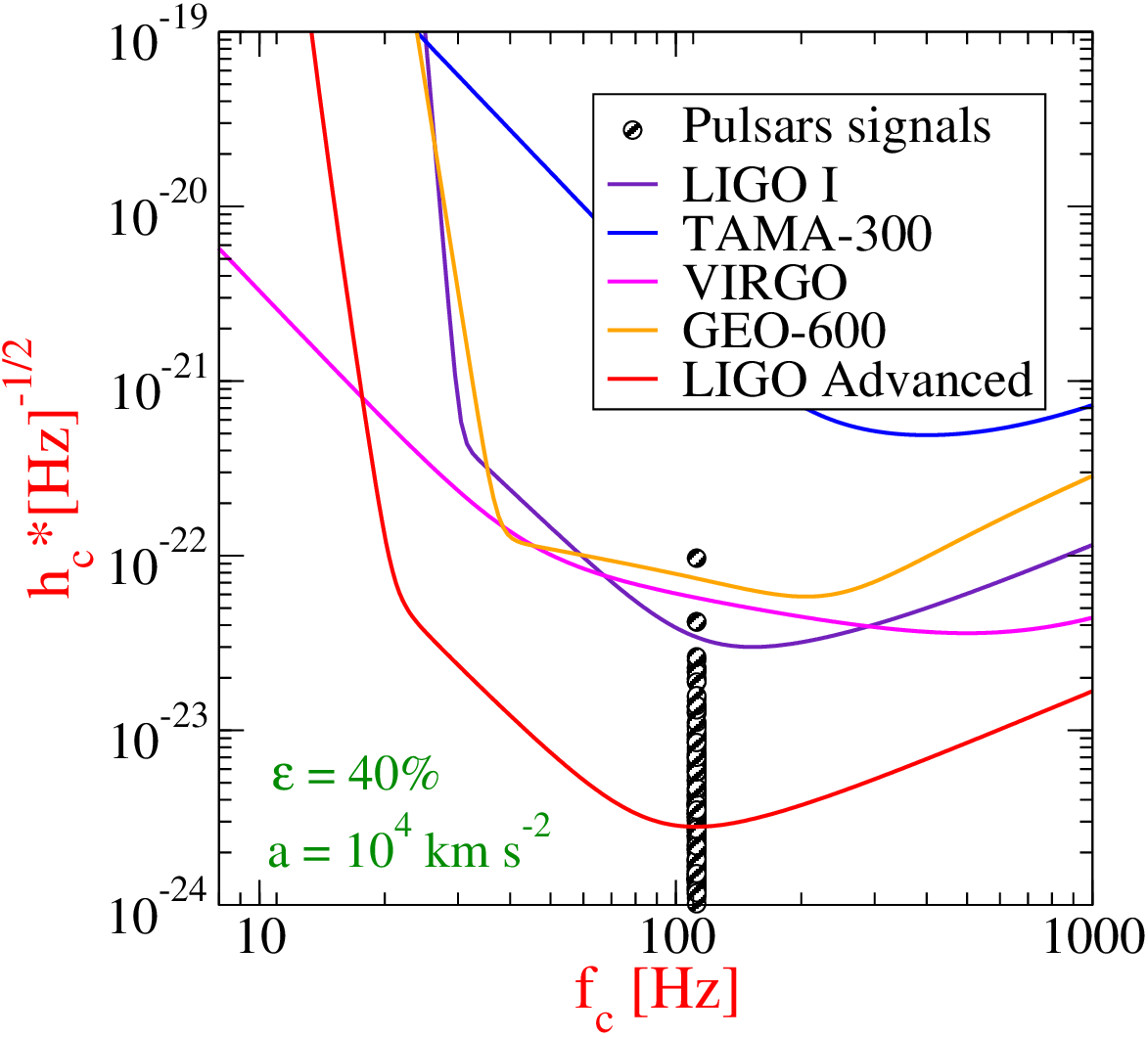} }
\hskip 0.5truecm
\subfigure[\label{L1-eff=0.1} Efficiency case $\epsilon = 0.10$!! for parameters  
$a = 10^4$ km s$^{-2}$ and $V_{\rm kick} \sim$ 10$^4$ km s$^{-1}$, calibrated against LIGO I. ]{
\includegraphics[height=3.0in,width=3.0in]{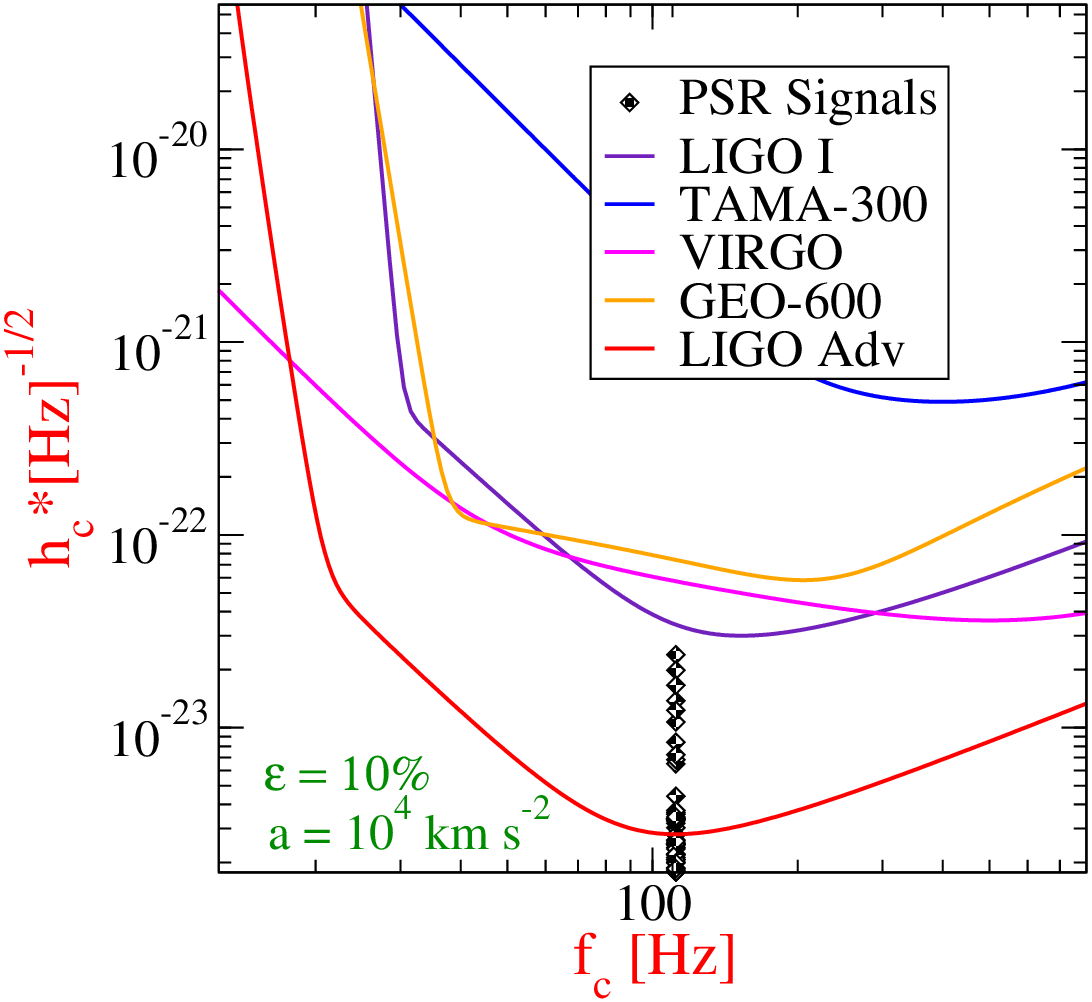} }
\vskip 1.0truecm
\subfigure[\label{L1-eff=0.05} Efficiency case $\epsilon = 0.05$!! for parameters  
$a = 10^4$ km s$^{-2}$ and $V_{\rm kick} \sim$ 10$^4$ km s$^{-1}$, calibrated 
against LIGO I. ]{
\includegraphics[height=3.0in,width=3.0in]{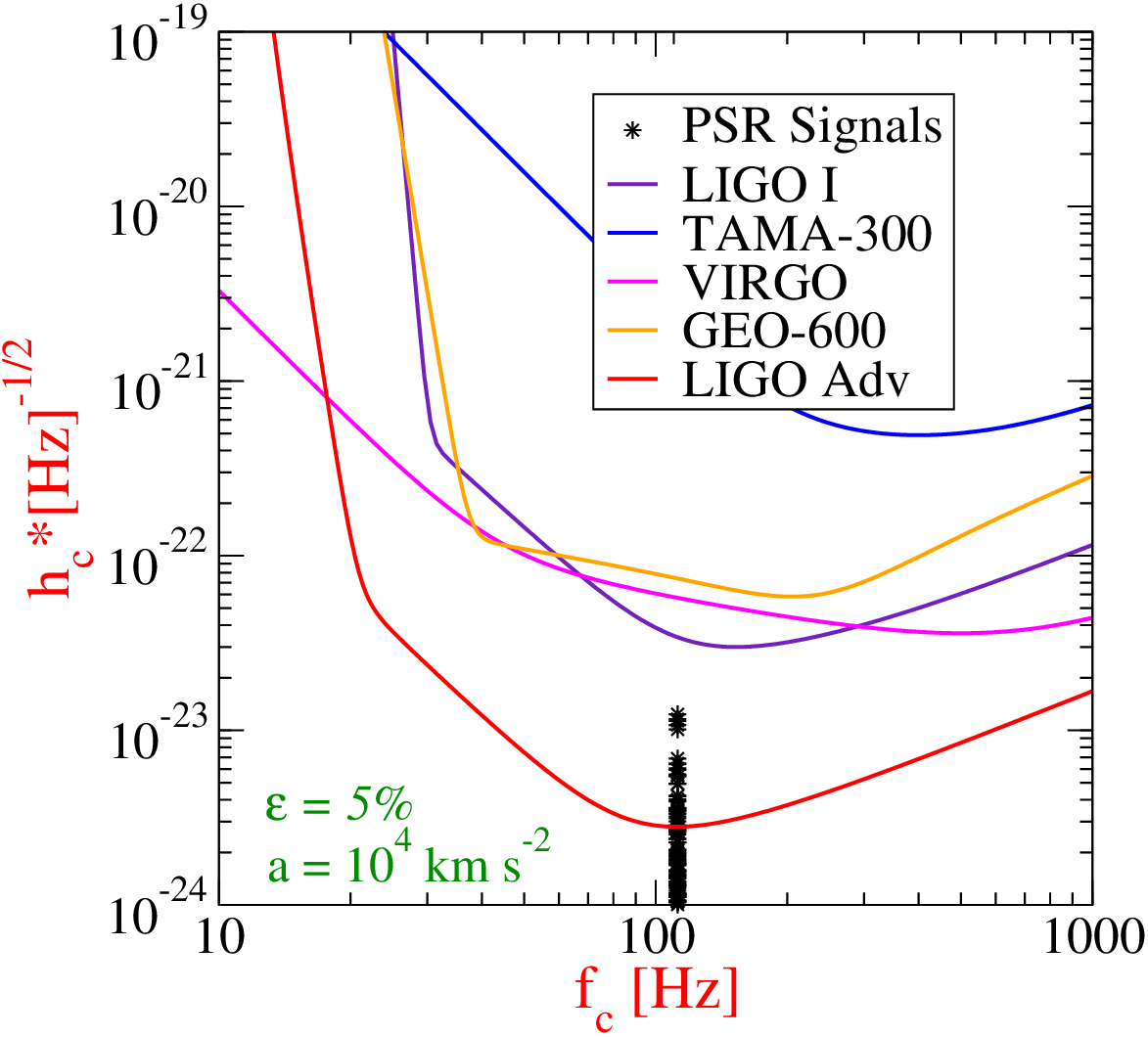} }
\hskip 0.5truecm
\subfigure[\label{L1-eff=0.01} Efficiency case $\epsilon = 0.01$!! for parameters  
$a = 10^4$ km s$^{-2}$ and $V_{\rm kick} \sim$ 10$^4$ km s$^{-1}$, calibrated 
against LIGO I. ]{
\includegraphics[height=3.0in,width=3.0in]{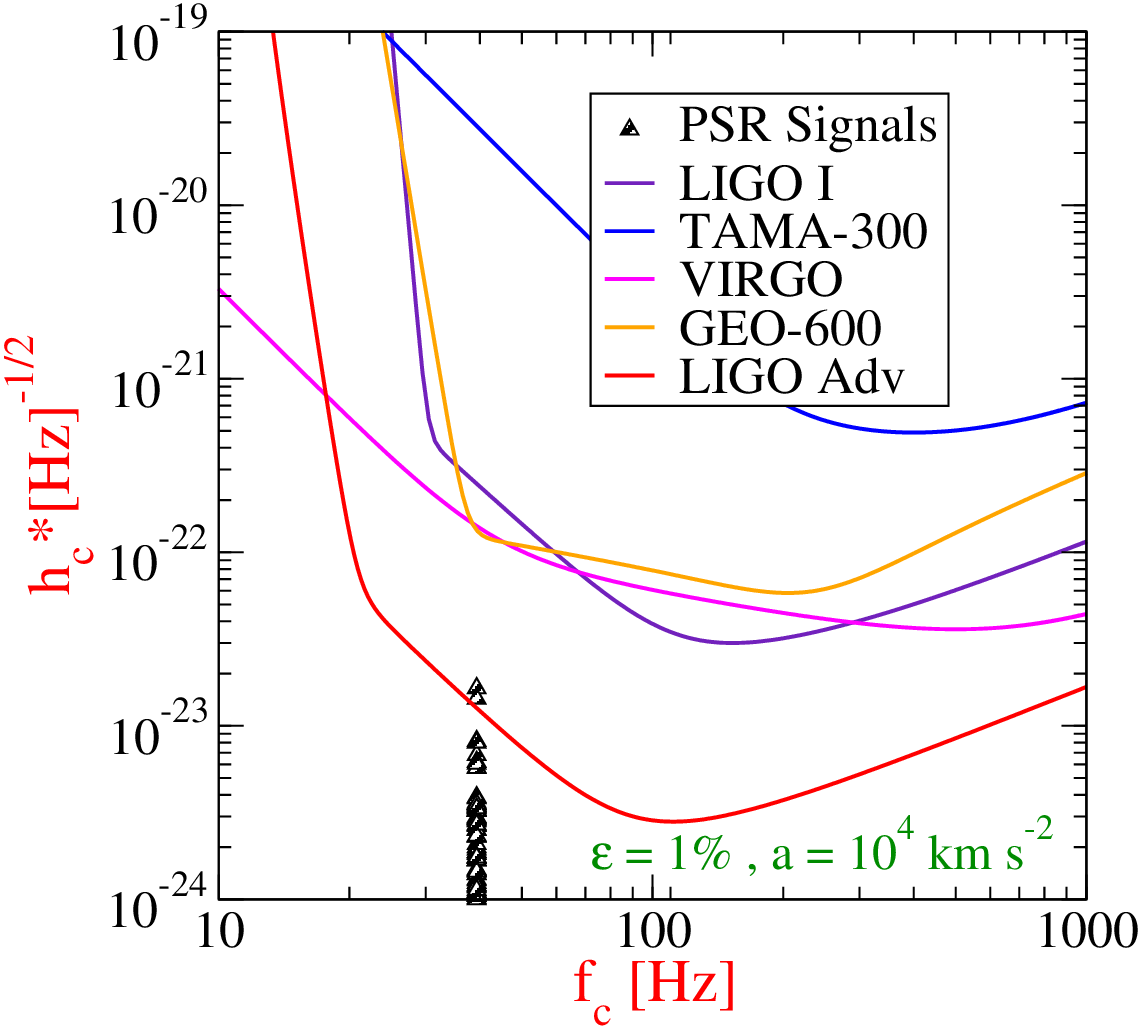} }
\caption{Pulsar GW signals for parameters: $a = 10^4$ km s$^{-2}$ and $V_{\rm kick} 
\sim$ 10$^4$ km s$^{-1}$}
\label{L1EFFICIENCY-a4}
\end{figure*}

\begin{figure*}[tbh]
\centering
\subfigure[\label{L1eff1=0.4a} Efficiency case $\epsilon = 0.40$!! for parameters  
$a = 10^6$ km s$^{-2}$ and $V_{\rm kick} \sim$ 10$^4$ km s$^{-1}$, calibrated 
against LIGO I. ]{
\includegraphics[height=3.0in,width=3.0in]{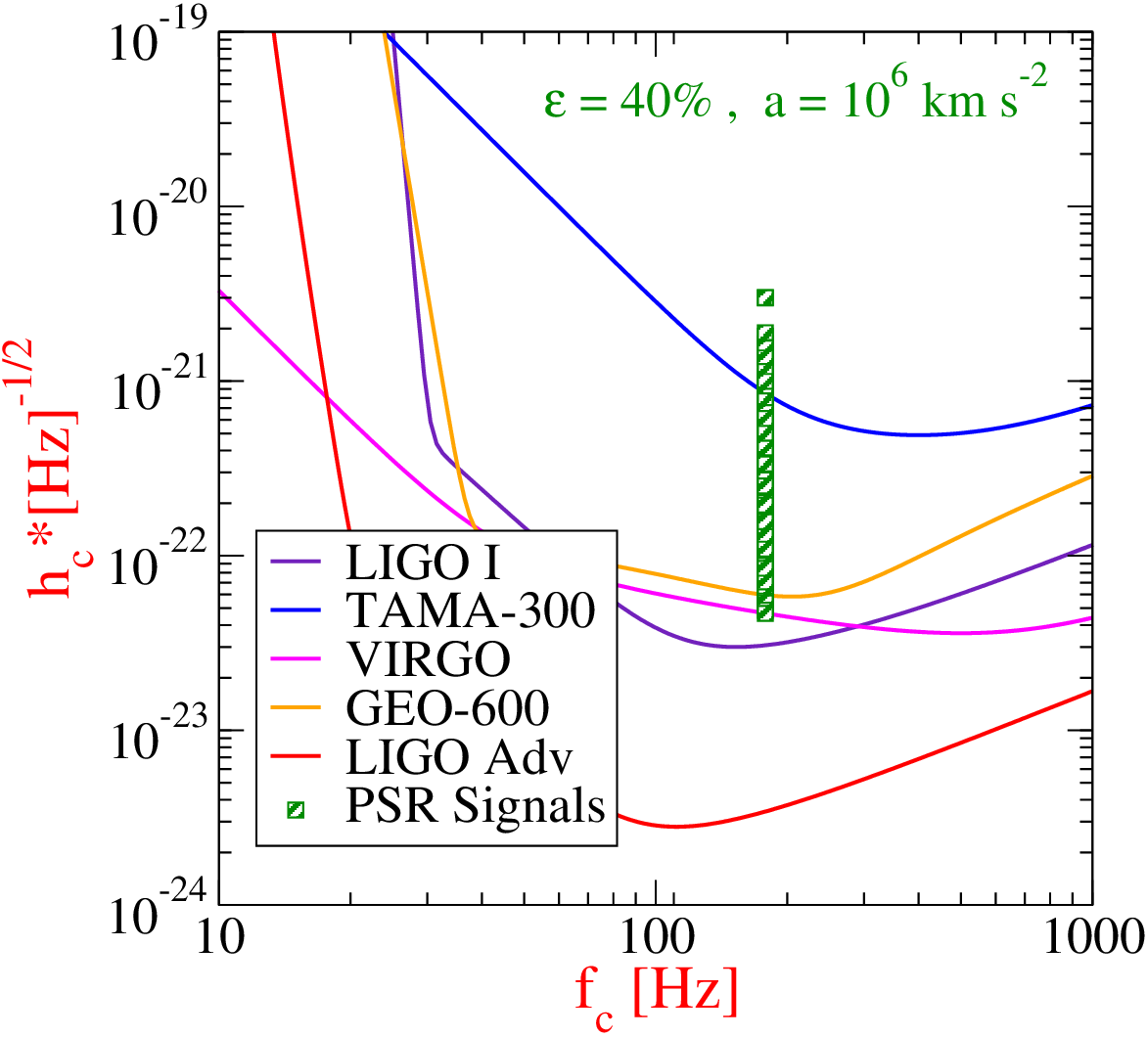} }
\hskip 0.5truecm
\subfigure[\label{L1eff1=0.4b} Efficiency case $\epsilon = 0.10$!! for parameters  
$a = 10^6$ km s$^{-2}$ and $V_{\rm kick} \sim$ 10$^4$ km s$^{-1}$, calibrated 
against LIGO I. ]{
\includegraphics[height=3.0in,width=3.0in]{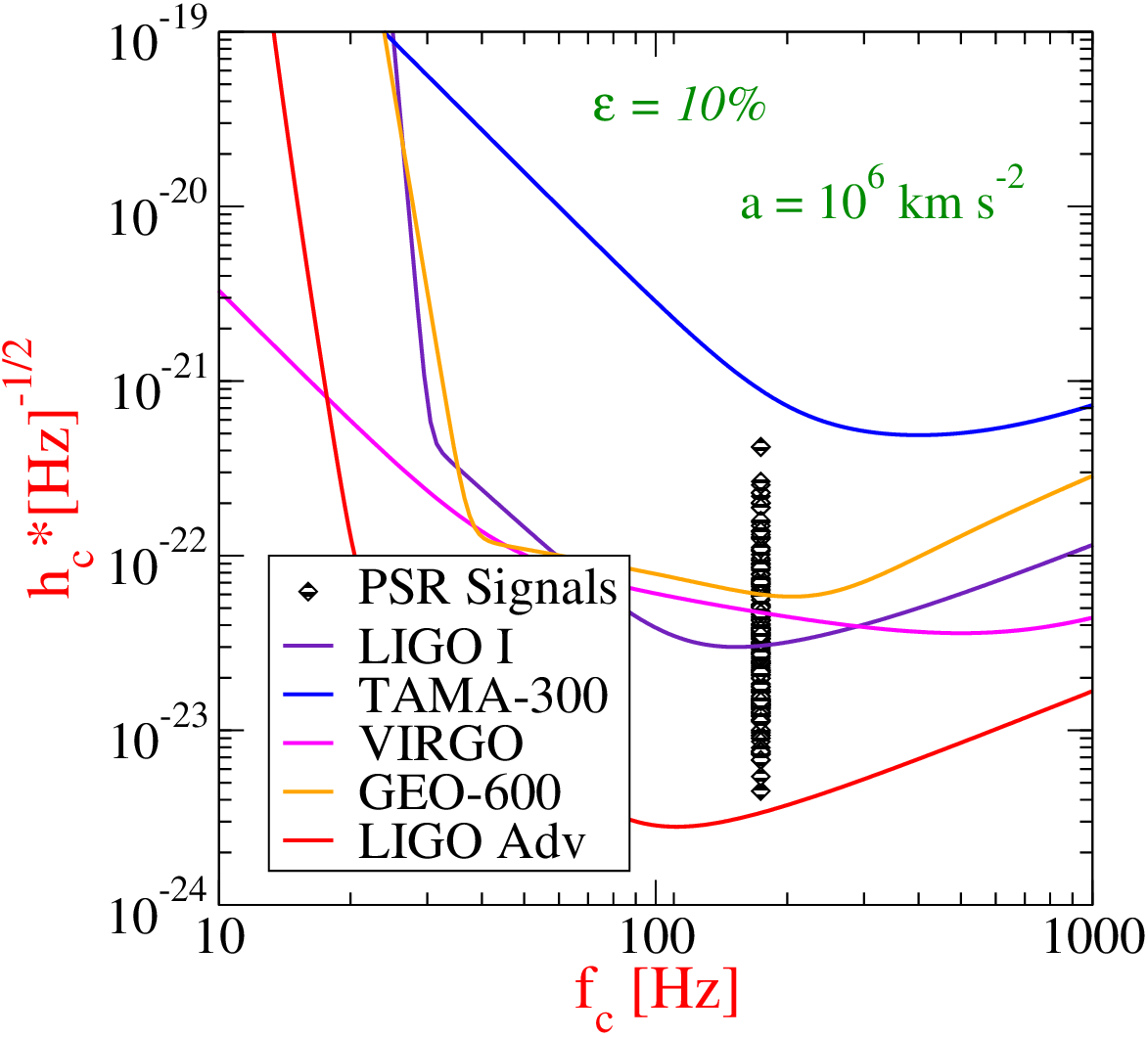} }
{\vskip 1.0truecm
\subfigure[\label{L1eff1=0.4c} Efficiency case $\epsilon = 0.05$!! for parameters  
$a = 10^6$ km s$^{-2}$ and $V_{\rm kick} \sim$ 10$^4$ km s$^{-1}$, calibrated 
against LIGO I. ]{
\includegraphics[height=3.0in,width=3.0in]{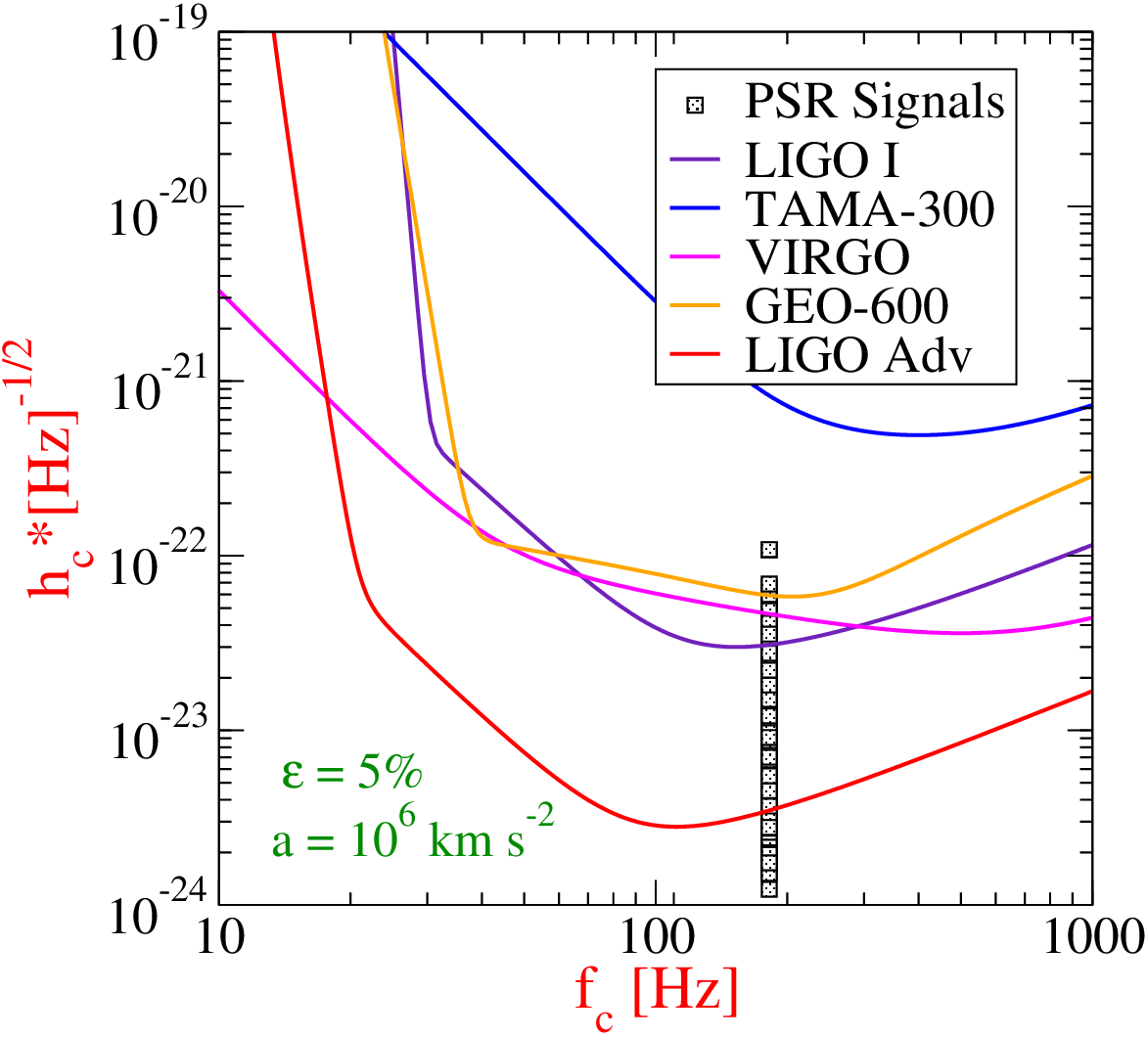} }
\hskip 0.5truecm
\subfigure[\label{L1eff1=0.4d} Efficiency case $\epsilon = 0.01$!! for parameters  
$a = 10^6$ km s$^{-2}$ and $V_{\rm kick} \sim$ 10$^4$ km s$^{-1}$, calibrated 
against LIGO I. ]{
\includegraphics[height=3.0in,width=3.0in]{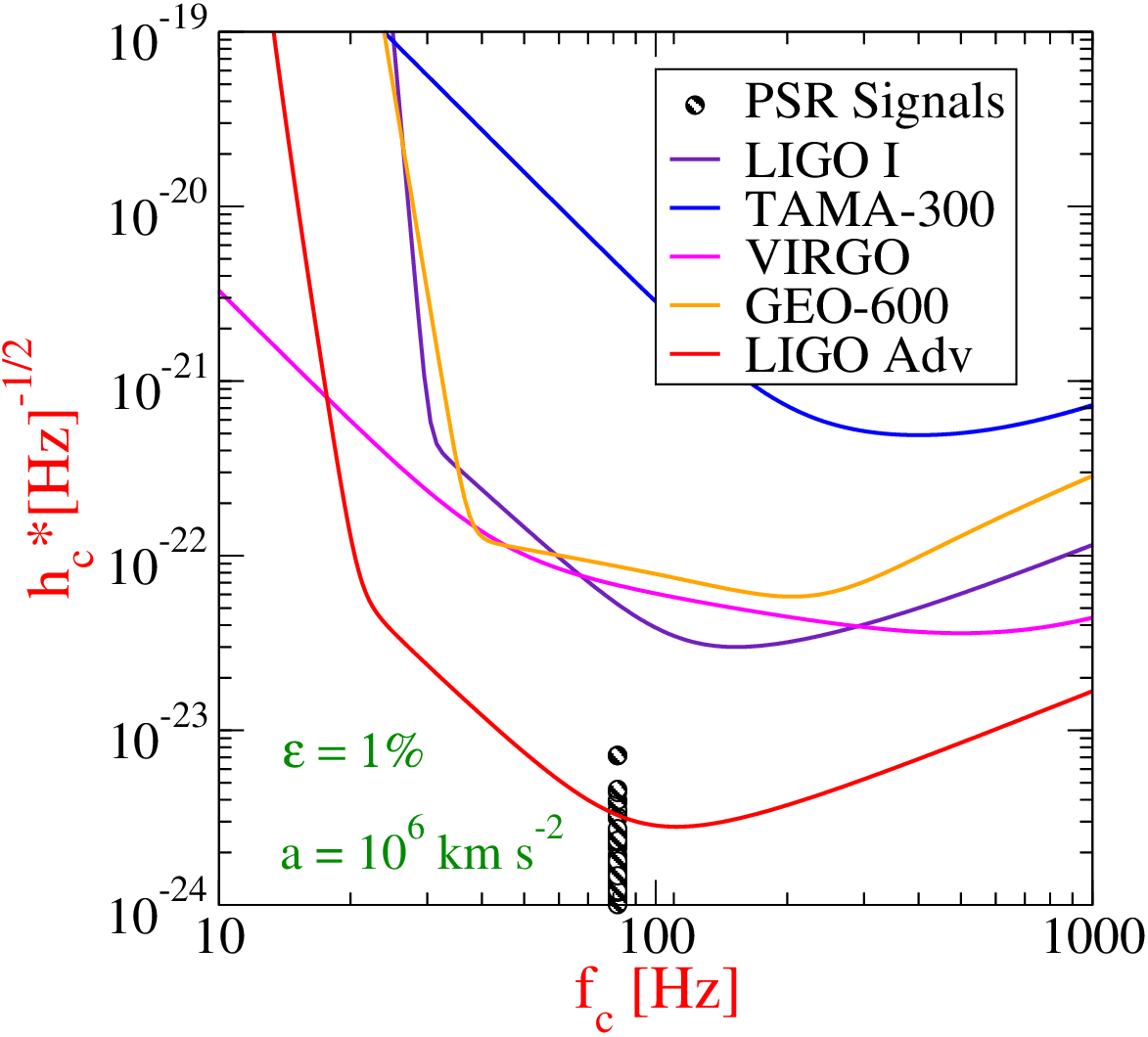} }
}
\caption{Pulsar GW signals for parameters: $a = 10^6$ km s$^{-2}$ and $V_{\rm kick} 
\sim$ 10$^4$ km s$^{-1}$}
\label{EFF=CALIB-LIGO-I}
\end{figure*}

The following figures illustrate the results obtained for this efficiency:
Fig.(\ref{L2eff1=0.4}) with $a = 10^6$ km s$^{-2}$ and Fig.(\ref{L2eff=0.4}) with 
$a = 10^4$ km s$^{-2}$ for Advanced LIGO. Fig.(\ref{L1eff1=0.4a}) with $a = 10^6$ km 
s$^{-2}$ and Fig.(\ref{L1-eff=0.4}) with $a = 10^4$ km s$^{-2}$ for LIGO I.

\section{Discussion and Conclusions}

\subsection{Computed rotation periods vs. Periods of ATNF sample of RAPs 
with $V=400-500$ km s$^{-1}$}

To the end, in order to make our analysis self-consistent we next discuss 
the relation between the translational speed and rotation frequency of ATNF 
pulsars. To do this, we select, as an example, the 3-D average pulsar velocity 
as determined by Hobbs et al. \cite{Hobbs2005} (see Table-\ref{periodos}).
On this basis we determine the theoretical rotation periods that one could 
expect to find for that velocity sample if is factual that most of the 
explosion energy, as inferred from the momentum conservation law in 
Eq.(\ref{momento}), goes into the pulsar rotation as discussed above. 
(Of course we compare with the actual periods measured by the ATNF 
surveys). Hence the torque applied to the just-born neutron star reads 

\begin{equation}
\tau \sim \frac {V}{2E_{k}} \frac{\frac{2}{5} M_{NS}R_{NS}^{2}}{d_{l}}
\end{equation}
\noindent

where $E_{k}$ is the kinetic energy of each velocity chosen for the present 
analysis, $V$ pulsar velocity, M$_{NS}=1.4$ M$_{\odot}$, R$_{NS} = 20$ km
and d$_{l} = 25$ km are the neutron star mass, radius and torque lever. 
Thus, for R$_{NS} = 20$ km and velocity $V=350$ km s$^{-1}$ we obtain p$ = 0.0182$ 
s, while for $V=433$ km s$^{-1}$ one gets p$=0.0182$ s. Meanwhile, for R$_{NS}=30$ 
km and $V=350$ km s$^{-1}$ we have p$=0.0411$ s, while for $V=433$ km s$^{-1}$ 
one gets p$=0.0332$ s. 

One can then compare these results with those presented in Table-\ref{periodos}. It 
appears that the theoretical prediction is about an order of magnitude smaller (much 
shorter periods) than the observed, and measured by the ATNF surveys. This discrepancy 
may suggest that 
\begin{itemize}
\item 1) perhaps a large part of the initial rotational energy is dissipated 
by processes as differential rotation or magnetic braking, especially if processes as 
the magneto-rotational instability\cite{ozernoy-somov-mri} indeed drives a large class 
of supernova explosions \cite{MAGN-ROTAT-INSTAB}. 
\item 2) it may also happen that convection and neutron fingers also becomes highly 
effective 
dissipative processes, thus reducing the pulsar initial spin to the level that we are 
currently finding in pulsar surveys.
\end{itemize}

In summary, from astronomical statistics the number of observed Galactic RAPs 
is definitely much larger than the corresponding to relativistic NS-NS binaries. 
The typical lifetime of canonical pulsars is $\sim 10$ Myr, which is defined 
as the duration of the radio emitting phase for objects with $B \sim 10^{12}$ G, 
or the time needed for the pulsar to emit less radiation or being entirely turned 
off. Hence, with a statistical Galactic rate of SNe, and the likely fling out 
rate of just-born pulsars, of about 1 per every 30-300 years, one expects that 
the GW signal of the kick at birth be observable as aftermath of the SN core 
collapse. In other words, the RAP GW signal may become a benchmark for looking 
for GW signals from the SN gravitational implosion. Besides, if the GW signal 
from the kick to a RAP were detected, for instance, from a source inside a 
globular, or from either an open star cluster \cite{pulsars-vlemmings2005}, 
or a binary system harboring a compact star, such an observation might turn 
these RAPs the most compelling evidence for the existence of Einstein's 
gravitational waves. 

We conclude by stating that the detection with advanced LIGO-type interferometers 
of individual GW signals from the short rise fling of Galactic RAPs is indeed possible.

\begin{table}
\caption{Selected sample from the ATNF catalog of pulsars having speeds between 
350-500 km s$^{-1}$, and their corresponding periods.}
\begin{center}
\begin{tabular}{llll}
\hline \hline
$\#$ & JNAME & PERIOD & VTRANS  \\
{ }&{ }& [s] & [km s$^{-1}$] \\
\hline\hline
1  & \text{J0255-5304}  &   0.447708  &  381.648\\
2  & \text{J0538+2817}  &   0.143158  &  401.682\\
3  & \text{J0543+2329}  &   0.245975  &  377.151\\
4  & \text{J0837-4135}  &   0.751624  &  364.772\\
5  & \text{J1300+1240}  &   0.006219  &  350.637\\
6  & \text{J1645-0317}  &   0.387690  &  417.022\\
7  & \text{J1803-2137}  &   0.133617  &  351.252\\
8  & \text{J1902+0615}  &   0.673500  &  368.086\\
9  & \text{J1941-2602}  &   0.402858  &  351.027\\
10 & \text{J2048-1616}  &   1.961572  &  355.328\\
11 & \text{J2113+2754}  &   1.202852  &  386.791\\
12 & \text{J2219+4754}  &   0.538469  &  375.304\\
13 & \text{J2305+3100}  &   1.575886  &  373.546\\
14 & \text{J2326+6113}  &   0.233652  &  433.576\\
15 & \text{J2354+6155}  &   0.944784  &  357.846\\
\hline \hline
\label{periodos}
\end{tabular}
\end{center}
\end{table}


\acknowledgments{HJMC thanks Prof. Jos\'e A. de Freitas Pacheco (OCA-Nice) for 
the hospitality during the initial development of this idea. Dr. Sebastien Peirani 
(OCA-Nice), Marcelo Perantonio (CBPF-Rio de Janeiro), and folks as Dr. Eric Lagadec 
(OCA-Nice) are also thanked. HJMC is fellow of Funda\c c\~ao de Amparo \`a Pesquisa 
do Estado do Rio de Janeiro (FAPERJ), Brazil.}


\begin{thebibliography}{99}

\bibitem{atnf-manchester2004}R. N. Manchester, G. B. Hobbs, A. Teoh, M. Hobbs, 
{\sl The ATNF pulsar catalogue} (Australia, CSIRO, Epping), Astron. J 129, 
1993-2006 (2005), e-Print: astro-ph/0412641 

\bibitem{lorimer93} Lorimer, D.R., et al., MNRAS 263, 403 (1993). 

\bibitem{lorimer94} Lyne, A.G. \& Lorimer, D.R., 1994, Nature 369, 127, 

\bibitem{arzoumian2002}Z. Arzoumian, D. Chernoff, J. M. Cordes, Astrophys. J. 
568, 289 (2002)

\bibitem{phinney-spruit98} Spruit, H. \& Phinney, S. E., Nature 393, 139 (1998)   . 


\bibitem{braginsky-thorne83} V. B. Braginsky, K. S. Thorne, Nature 327., 
123-125 (1987)

\bibitem{cordes-chernoff98}J. M. Cordes, D. F. Chernoff,  Astrophys. J. 505, 
315-338 (1998) 


\bibitem{chakra-thorsett} D. Chakrabarty, S. E. Thorsett, Astrophys. J. 512, 288-299 (1998) 

\bibitem{PMBS}R. N. Manchester, G. B. Hobbs, A. Teoh, M. Hobbs, `The ATNF 
Pulsar Catalog'. Report astro-ph/0412641 (2004). The ATNF web-site is 
found via the URL link http://www.atnf.csiro.au/research/pulsar/psrcat 

\bibitem{Hobbs2005} G. Hobbs et al., MNRAS 360, 974 (2005) 

\bibitem{Abbott2007}B. Abbott et al., LIGO Science Collaboration, arXiv:0704.3368L
\bibitem{postnov2006}K. A. Postnov, L. Youngelson, Liv. Rev. Rel. 9, 6 (2006)

\bibitem{kalogera2007}V. Kalogera et al. Phys. Rep. 442, 75 (2007)

\bibitem{taylor2004} Weisberg, J. M. Taylor, J. H., 2003. The
Relativistic Binary Pulsar B1913+16. Conference Proceedings, Vol. 302.
Held 26-29 August 2002 at Mediterranean Agronomic Institute of Chania,
Crete, Greece. Edited by Matthew Bailes, David J. Nice and Stephen E.
Thorsett. San Francisco: Astronomical Society of the Pacific, 2003.
ISBN:  1-58381-151-6, p.93

\bibitem{Wijnands1998} R. Wijnands and M. van der Klis, Nature 394, 344 
(1998)

\bibitem{pulsars-vlemmings2005}W.H.T. Vlemmings, S. Chatterjee, W.F. Brisken, 
T.J.W. Lazio, J.M. Cordes, S.E. Thorsett, W.M. Goss, E.B. Fomalont, M. Kramer, 
A.G. Lyne, S. Seagroves, J.M. Benson, M.M. McKinnon, D.C. Backer, R. Dewey, 
{\sl Pulsar Astrometry at the Microarcsecond Level}. In the proceedings of the 
"Stellar End Products" workshop, 13-15 April 2005, Granada, Spain (for 
publication in MmSAI vol.77).  arXiv:astro-ph/0509025 

\bibitem{pulsars-gaia} Coryn A. L. Bailer-Jones 
{\sl  Microarcsecond astrometry with Gaia: The Solar system, the Galaxy and beyond}. 
Invited paper at IAU Colloquium 196: Transits of Venus: New Views of the Solar System 
and Galaxy, Lancashire, England, United Kingdom, 7-11 Jun (2004) Submitted to IAU Symp.
e-Print: astro-ph/0409531. See also O. Fors in {\sl New observational techniques 
and analysis tools for wide field CCD surveys and high resolution astrometry}
(2006). e-Print: astro-ph/0604150. Also see E. Fomalont, M. Reid, {\sl Microarcsecond 
Astrometry using the SKA}, New Astron. Rev. 48, 1473-1482 (2004) 

\bibitem{sago04}N. Sago, et al., Phys. Rev. D 70, 104012-1,104012-8 (2004). 

\bibitem{ori2001}B. Segalis, A. Ori, Phys. Rev. D 64, 064018 (2001) 


\bibitem{pulsars-background}H. J. Mosquera Cuesta and C. A. Bonilla Quintero, 
{\sl The gravitational-wave background produced by the kick at birth of galactic 
run away pulsars}, (accompanying paper in preparation) (2007)

\bibitem{pacheco2006} Jose A. de Freitas Pacheco, Tania Regimbau, S. Vincent, A. 
Spallicci Int. J. Mod. Phys. D15:235-250 (2006).e-Print: astro-ph/0510727: {\sl 
Expected coalescence rates of ns-ns binaries for laser beam interferometers}

\bibitem{MAGN-ROTAT-INSTAB}A. Ud-Doula, J. Blondin, {\sl The Magneto-Rotational 
Instability in Core-Collapse Supernovae}, American Physical Society, The 70th Annual 
Meeting of the Southeastern Section, November 6-8, 2003, Wilmington, North Carolina, 
MEETING ID: SES03, abstract \#EB.009. See also S. Akiyama, J. C. Wheeler, D. L. Meier,  
and I. Lichtenstadt, AAS 199.9403 (2001); J. Craig Wheeler, {\sl Rotation and Magnetic 
Fields in Supernovae and Gamma-ray Bursts}, American Physical Society, 47th Annual DPP 
Meeting, October 24-28 (2005), abstract \#CZ2.002; P. Cerdá-Durán, J. A. Font, H. Dimmelmeier, 
Astron. \& Astrophys. 474, 169 (2007); L. Dessart, A. Burrows, E. Livne, C. Ott, {\sl The 
Proto-neutron Star Phase of the Collapsar Model and the Route to Long-soft Gamma-ray Bursts 
and Hypernovae},  arXiv: 0710.5789 (2007); M. Shibata, Y. T. Liu, S. L. Shapiro, C. B. 
Stephens, PRD 74, 104026 (2006), {\sl Magnetorotational collapse of massive 
stellar cores to neutron stars: Simulations in full general relativity}

\bibitem{fryer2006} Chris L. Fryer, Astrophys.J.601:L175-L178,2004.
e-Print: astro-ph/0312265, {\sl Neutron star kicks from asymmetric collapse.}. See also, 
Aristotle Socrates, Omer Blaes, Aimee L. Hungerford, Chris L. Fryer Astrophys. J. 632:
531-562 (2005). e-Print: astro-ph/0412144, {\sl The Neutrino bubble instability: A 
Mechanism for generating pulsar kicks }; Chris L. Fryer, Alexander Kusenko, Astrophys. 
J. Suppl. 163:335 (2006), e-Print: astro-ph/0512033, {\sl Effects of neutrino-driven 
kicks on the supernova explosion mechanism}

\bibitem{janka-muller-review}Leonhard Scheck, K. Kifonidis, H.-Th. Janka, E. Mueller, 
{\sl Multidimensional supernova simulations with approximative neutrino transport. 
1. neutron star kicks and the anisotropy of neutrino-driven explosions in two 
spatial dimensions}, Submitted to Astron.Astrophys. e-Print: astro-ph/0601302. 
See also Hans-Thomas Janka, L. Scheck, K. Kifonidis, E. Muller, T. Plewa
{\sl Supernova asymmetries and pulsar kicks - Views on controversial issues}.
e-Print: astro-ph/0408439

\bibitem{dl2004}D. Lai (2004). In {\sl cosmic explosions in three dimensions: 
asymmetries in supernovae and gamma-ray bursts}, eds. P. Hoflich, P. Kumar, J. 
C. Wheeler (Cambridge Univerity Press, Cambridge), p. 276

\bibitem{dong-lai2001} Dong Lai, {\sl Neutron star kicks and asymmetric supernovae}. 
e-Print: astro-ph/0012049. See also Chen Wang, Dong Lai, JinLin Han, Astrophys.J.639,
1007-1017 (2006). e-Print: astro-ph/0509484: {\sl Neutron star kicks in isolated and 
binary pulsars: observational constraints and implications for kick mechanisms}

\bibitem{ozernoy-somov-mri}L. M. Ozernoy, B. V. Somov,Ap. S S 11, 264O(1971)
{\sl The Magnetic Field of a Rotating Cloud and Magneto-Rotational Explosions}

\end{thebibliography}
\end{document}